\documentstyle[12pt,epsfig,graphics,amssymb]{article}
\begin{document}
\setlength{\parskip}{0.45cm}
\setlength{\baselineskip}{0.75cm}
%XXXXXXXXXXXXXXXXXXXXXXXXXXXXXXXXXXXXXX
%
%SETTINGS FOR PREPRINT-SPACED VERSION
%setlength{\parskip}{0.45cm}
%setlength{\baselineskip}{0.75cm}
%
% SETTINGS FOR DOUBLE - SPACED VERSION
%\setlength{\parskip}{0.65cm}
%\setlength{\baselineskip}{0.95cm}
%
%XXXXXXXXXXXXXXXXXXXXXXXXXXXXXXXXXXXXXX
\begin{titlepage}
\setlength{\parskip}{0.25cm}
\setlength{\baselineskip}{0.25cm}
\begin{flushright}
DO-TH 2000/14\\
\vspace{0.2cm}
RBRC-148 \\
\vspace{0.2cm}
TPR-00-21\\
\vspace{0.2cm}
hep--ph/0011215\\
\vspace{0.2cm}
October 2000
\end{flushright}
\vspace{0.6cm}
\begin{center}
\LARGE
{\bf Models for the Polarized Parton \\
\vspace{0.1cm}
 Distributions of the Nucleon}
\vspace{1.2cm}

\large
M. Gl\"uck$^a$, E.\ Reya$^a$, M.\ Stratmann$^b$, W.\ Vogelsang$^c$\\
\vspace{1.0cm}

\normalsize
$^a${\it Universit\"{a}t Dortmund, Institut f\"{u}r Physik,}\\
{\it D-44221 Dortmund, Germany} \\
\vspace{0.5cm}

$^b${\it Institut f\"ur Theoretische Physik, Universit\"at Regensburg}\\
{\it D-93040 Regensburg, Germany}\\
\vspace{0.5cm}

$^c${\it RIKEN-BNL Research Center, Brookhaven National Laboratory\\
{\it Upton, NY 11973, USA}}\\

\vspace{1.5cm}
\end{center}

\begin{abstract}
\noindent
Polarized deep inelastic scattering (DIS) data are analyzed
in leading and next--to--leading order of QCD within the common
 `standard' scenario of polarized parton distributions with a
flavor--symmetric light sea (antiquark) distribution $\delta\bar{q}$,
and a completely SU(3)$_f$ broken `valence' scenario with totally
flavor-asymmetric
light sea densities $(\delta\bar{u}\neq\delta\bar{d}\neq\delta\bar{s})$. 
The latter flavor--broken light sea distributions are modelled with 
the help of a Pauli--blocking ansatz at the low radiative/dynamical
input scales of $\mu_{\rm LO(NLO)}^2=0.26$ (0.40) GeV$^2$ which 
complies with predictions of the chiral quark--soliton model and
expectations based on the statistical parton model as well as with 
the corresponding, well established, flavor--broken
unpolarized sea ($\bar{d}>\bar{u}$).  Present  
semi--inclusive DIS data cannot yet uniquely discriminate between
those two flavor--symmetric and flavor--broken polarized light sea
scenarios.
\end{abstract}
\end{titlepage}

%MAIN PART

\renewcommand{\theequation}{\arabic{section}.\arabic{equation}}
\section{Introduction}

The polarized parton distributions of the nucleon have been
intensively studied in recent years [1 -- 14]. 
%\cite{ref1}--\cite{ref14}.
The conclusion has been that the experimental data dictate a
negatively polarized antiquark component, and show a tendency toward
a positive polarization of gluons. Presently we possess a lot
of precise data [15 -- 24]  
%\cite{ref15}--\cite{ref24}
on the polarized
structure functions of the nucleon, some of them very recent 
\cite{ref23,ref24}, which justify a renewed
investigation of the aforementioned issue.  This alone, however,
does not provide the main motivation for this project, rather the
improved understanding in recent years of the situation in the
{\em{unpolarized}} parton sector [25 -- 28] 
%\cite{ref25}--\cite{ref28} 
provides important insights for the
corresponding {\em{polarized}} parton densities.  In
particular, one notes that the unpolarized sea (antiquark)
distributions are flavor--asymmetric ($\bar{d}>\bar{u}$), which
can be understood in terms of flavor mass asymmetries and
Pauli--blocking effects \cite{ref29,ref30}. The main objective of
the present paper is to transcribe these insights into the
polarized parton sector as will be described in Section 3. In
Section 2 we shall, for completeness, present an analysis within
the framework of the simplified SU$(3)_f$ symmetric `standard'
scenario in which the flavor--asymmetries in the polarized
antiquark sector are neglected.  This is done in view of the fact
that in many situations these flavor asymmetries are unobservable
as is the case for (most) presently available data which cannot
provide any reliable information concerning this issue.

Measurements of polarized deep--inelastic lepton nucleon
scattering yield direct information [15 -- 24]
%\cite{ref15}--\cite{ref24}
 on
the spin asymmetry
%Eq. (1.1)
\begin{equation}
A_1^N(x,Q^2)\simeq \frac{g_1^N(x,Q^2)}{F_1^N(x,Q^2)} =
   \frac{g_1^N(x,Q^2)}{F_2^N(x,Q^2)/\{ 2x[1+R^N(x,Q^2)]\}}\, ,
\end{equation}
$N=p,\,n$ and $d=(p+n)/2$ where in the latter case we have used
$g_1^d=(g_1^p+g_1^n)[1-3\omega_D/2]/2$ with $\omega_D=0.058$,
$R\equiv F_L/2xF_1=(F_2-2xF_1)/2xF_1$ and subdominant
contributions have, as usual, been neglected.

We emphasize that, as in our original analysis \cite{ref1}, we
compute {\it both} $g_1$ and $F_1$ entirely in leading--twist
QCD.  In particular, in order to obtain $F_1$, we use the parton
densities of GRV98 \cite{ref25} along with LO (note that $R=0$ at
leading order) or NLO coefficient functions for $F_2$ and $R$ in
(1.1). An alternative, frequently adopted
\cite{ref3,ref4,ref5,ref9,ref10,ref13} approach is to take $F_2(x,Q^2)$
and $R(x,Q^2)$ from experimental measurements, which is motivated
by the fact that leading--twist calculations of $F_2(x,Q^2)$ and
$R(x,Q^2)$ do not agree very well with experimental
determinations in the region of low $Q^2$ and $W^2=Q^2(1-x)/x$.
These regions are affected by power--suppressed contributions and
are therefore excluded from all unpolarized DIS analyses.  The
presently available data in the polarized case, however, do not
allow to impose similarly `safe' cuts ($Q^2\geq 4$ GeV$^2$,
$W^2\geq10$ GeV$^2$) without losing too much information.
On the other hand, the $Q^2$ range accessed so far in polarized
DIS does not allow for extracting the magnitude and shape of
power--suppressed contributions reliably.  To study the issue
further, we performed fits to the $A_1$ data in both possible
ways, i.e.\ with leading--twist calculations of $F_2$ and $R$
{\it as well as} with their experimental results, admitting at
the same time an `effective higher--twist' contribution to $A_1$
in terms of a factor $(1+A(x)/Q^2)$ with $A(x)$ to be determined
by the data. The outcome of this analysis was that $A(x)$ is
consistent with zero if we use leading--twist QCD for $F_2$ and
$R$, but that it is sizeable and important in the fit if $F_2$
and $R$ are taken from experiment.  We take this as an indication
that our preferred approach is more consistent and less liable to
modifications by higher--twist terms. This view is also corroborated
by the fact that the DIS $A_1$ data show only a very mild
$Q^2$--dependence, even toward low values of $Q^2$.  
The consistency of the
polarized parton densities as extracted from DIS data at
comparatively low values of $Q^2$ with measurements of other hard
processes at higher scales can be studied soon at RHIC and perhaps
in the future at a polarized {\it ep} collider.

 In NLO, $g_1^N(x,Q^2)$ is related
to the polarized (anti)quark and gluon distributions $\delta f
(x,Q^2)\equiv f_+ -f_-$ in the following way:
%Eq.(1.2)
\begin{equation}
g_1^N(x,Q^2) = \frac{1}{2} \sum_q e_q^2\{\delta q^N(x,Q^2) +
   \delta\bar{q}\,^N(x,Q^2) + \frac{ \alpha_s(Q^2)}{2\pi}
    [\delta C_q*(\delta q^N+\delta \bar{q}^N) + \frac{1}{f}
       \delta C_g*\delta g]\}
\end{equation}
with the convolutions $\delta C_f*f$ being defined in the usual
way. The $\overline{\rm{MS}}$ coefficient functions $\delta
C_f(x)$ can be also found in \cite{ref1}, where all necessary
ingredients for the $Q^2$--evolution have been formulated as
well.  A similar expression holds for the unpolarized structure
function $F_1^N(x,Q^2)$ with its spin--averaged parton distributions
$f(x,Q^2)\equiv f_+ +f_-$ and the unpolarized Wilson coefficients
can be found, for example, in \cite{ref31}.  The LO and
NLO($\overline{\rm{MS}}$) input scales, running coupling
constants and  parton distributions, employed in the positivity
constraints $|\delta f|\leq f$, will be adopted from GRV98
\cite{ref25}. Furthermore we shall, as always, use the notation
$\delta q^p\equiv\delta q$ and $q^p\equiv q$, and neglect the
marginal charm contribution to $g_1^N$ stemming from the
subprocess $\vec{\gamma}\,^*\vec{g}\to c\bar{c}$ \cite{ref32}. The
charm contribution to $F_1^N$ is also small in the kinematic
range covered by present polarization experiments.

The total helicity of a specific parton $f=u,\, \bar{u},\, d,\, \bar{d},\,
s,\, \bar{s},\, g$ is given by the first ($n=1$) moment
%Eq.(1.3)
\begin{equation}
\Delta f(Q^2) \equiv \int_0^1 dx\, \delta f(x,Q^2)\, .
\end{equation}
Thus, according to (1.2),
%Eq.(1.4)
\begin{equation}
\Gamma_1(Q^2) \equiv \int_0^1 dx\, g_1(x,Q^2) = \frac{1}{2}\sum_q e_q^2
    \left[ \Delta q(Q^2)+\Delta \bar{q}(Q^2)\right] \,
      \left( 1-\frac{\alpha_s(Q^2)}{\pi}\right)
\end{equation}
since $\Delta C_q=-3 C_F/2 = -2$ and $\Delta C_g=0$.  Therefore we
have in general
%Eq.(1.5)
\begin{equation}
\Gamma_1^{p,n}(Q^2) = \left[ \pm \frac{1}{12} \Delta q_3 + \frac{1}{36}
   \Delta q_8 + \frac{1}{9} \Delta\Sigma (Q^2)\right]\,
    \left( 1 - \frac{\alpha_s(Q^2)}{\pi}\right)
\end{equation}
with the flavor-nonsinglet components
%Eq.(1.6) u. (1.7)
\begin{eqnarray}
\Delta q_3 & \equiv & \Delta u+\Delta\bar{u} -\Delta d-\Delta\bar{d}\\
\Delta q_8 & \equiv & \Delta u+\Delta\bar{u} +\Delta d+\Delta\bar{d}-2
     (\Delta s+\Delta\bar{s})
\end{eqnarray}
being conserved, i.e.\ $Q^2$--independent,
and the flavor--singlet component is given by
%Eq.(1.8)
\begin{equation}
\Delta\Sigma(Q^2) \equiv \sum_{q=u,d,s}
   \left[ \Delta q(Q^2) + \Delta\bar{q} (Q^2)\right] =
     \Delta q_8 +3\left[\Delta s(Q^2)+\Delta\bar{s}(Q^2)\right]\, .
\end{equation}
These quantities will be subject to various constraints (derived from
hyperon $\beta$--decays) depending on the specific model scenarios
under consideration to which we shall turn in the next two Sections.

Finally, the fundamental helicity sum rule reads
%Eq.(1.9)
\begin{equation}
\frac{1}{2} = \frac{1}{2} \Delta\Sigma(Q^2) + \Delta g(Q^2) +
   L_{q+g}(Q^2)
\end{equation}
where $L_{q+g}$ refers to the total orbital contribution of all
(anti)quarks and gluons to the spin of the proton.

\setcounter{equation}{0}

\section{SU(3)$_f$ symmetric `standard' (unbroken--sea) scenario}
As stated in the Introduction, present data do not provide
sufficient information concerning the flavor--asymmetries of the
polarized sea distributions.  Thus present day studies must, as a
first approximation, neglect this issue unless one is willing to
adopt some \mbox{models} for the flavor--asymmetries as will be
done in Section 3.  Here we follow the procedure presented in
GRSV95 \cite{ref1}.  The searched for polarized NLO (as well as LO) parton
distributions $\delta f(x,Q^2)$, compatible with present data [15 -- 24]
%\cite{ref15}--\cite{ref24}
 on $A_1^N(x,Q^2)$, are constrained by
the `standard' sum rules
%Eq.(2.1)+(2.2)
\begin{eqnarray}
\Delta q_3 & = & F+D = g_A = 1.2670\,\, (35)\\
\Delta q_8 & = & 3F-D = 0.58\pm 0.15
\end{eqnarray}
where the updated values for $F$ and $D$ have been taken into
account \cite{ref33} and the error estimate in Eq.\ (2.2) is due
to \cite{ref34}.  Thus, Eq.\ (1.5) becomes
%Eq.(2.3)
\begin{equation}
\Gamma_1^{p,n}(Q^2) = \left[ \pm \frac{1}{12} (F+D) + \frac{5}{36} (3F-D)
  + \frac{1}{3} \left( \Delta s(Q^2)+\Delta\bar{s}(Q^2)\right) \right]
    \left(1-\frac{\alpha_s(Q^2)}{\pi}\right)\, ,
\end{equation}
i.e.\ one needs here a finite sizeable strange sea polarization
$\Delta s(Q^2)<0$ in order to achieve the experimentally required
reduction of the Ellis--Jaffe LO expectation \cite{ref35}
%Eq.(2.4)
\begin{equation}
\Gamma_{1,\rm{EJ}}^p = \frac{1}{12}(F+D) + \frac{5}{36}(3F-D)
   \simeq 0.186\, .
\end{equation}
Furthermore, in the `standard' scenario one assumes an unbroken
SU(3)$_f$ symmetric sea,
%Eq.(2.5)
\begin{equation}
\delta\bar{q}(x,Q^2) \equiv \delta\bar{u} = \delta u_{\rm{sea}} =
   \delta\bar{d} = \delta d_{\rm{sea}} = \delta s = \delta\bar{s}\, .
\end{equation}
For the determination of the NLO (LO) polarized parton
distributions $\delta f(x,Q^2)$ we follow our original analysis
\cite{ref1} by relating the polarized input densities to the unpolarized
ones, using some intuitive theoretical arguments \cite{ref36} as
guidelines. We employ
the following ansatz for the LO and NLO($\overline{\rm{MS}}$)
polarized parton distributions at an input scale $Q^2=\mu^2$
\cite{ref37}:
\vspace{-0.5cm}
%Eq.(2.6)
\begin{eqnarray}
\delta u(x,\mu^2) & = & N_u\, x^{\alpha_u} (1-x)^{\beta_u} \, u(x,\mu^2)_{\rm{GRV}}\nonumber\\
\delta d(x,\mu^2) & = & N_d\, x^{\alpha_d} (1-x)^{\beta_d} \, d(x,\mu^2)_{\rm{GRV}}\nonumber\\
\delta\bar{q}(x,\mu^2) & = & N_{\bar{q}}\,
x^{\alpha_{\bar{q}}}(1-x)^{\beta_{\bar{q}}}  \, \bar{q}
  (x,\mu^2)_{\rm{GRV}}\nonumber\\
\delta g(x,\mu^2) & = & N_g\, x^{\alpha_g}(1-x)^{\beta_g} \,
g(x,\mu^2)_{\rm{GRV}}
\end{eqnarray}
with the LO and NLO unpolarized input densities referring to the
ones of GRV98 \cite{ref25} and $\bar{q}\equiv(\bar{u}+\bar{d})/2$
should be considered as the reference light sea distribution for
the `standard' unbroken--sea scenario in (2.5). The parameters of
our optimal LO densities at $\mu_{\rm LO}^2=0.26$ GeV$^2$ and the
ones of the NLO$(\overline{\rm MS})$ densities at $\mu_{\rm
NLO}^2=0.40$ GeV$^2$ are given in Table I.  These optimal LO and
NLO$(\overline{\rm MS})$ fits correspond to a $\chi^2$ per degree
of freedom ($\chi_{\rm DF}^2$) of $\chi_{\rm DF,\, LO}^2=0.84$
and to $\chi_{\rm DF,\, NLO}^2=0.81$.  The polarized gluon density 
in (2.6) is,
as usual, rather weakly constrained by present data.  Although a
fully saturated (via the positivity constraint) gluon input
$\delta g(x,\mu^2)=\pm g(x,\mu^2)$ is disfavored, a less saturated
$\delta g(x,\mu^2)=\pm xg(x,\mu^2)$ input or even a vanishing (purely
dynamical) input $\delta g(x,\mu^2)=0$ are fully compatible with
present data.  The latter choice, however, seems to be unlikely
in view of $\delta\bar{q}(x,\mu^2)\neq 0$.

In Fig.\ 1 our NLO results are compared with the data on
$A_1^N(x,Q^2)$ as well as with our old original NLO($\overline{
\rm MS}$) fit \cite{ref1}.  The differences between these two
results are small, except perhaps for $A_1^n$ in the large-$x$
region.  Our new LO fit is similar to the NLO one shown in Fig.\ 1 by
the solid curves.  The $Q^2$--dependence of our LO and NLO `standard'
scenario fits at various fixed values of $x$ is shown in Fig.\ 2 and
compared with all recent data on $g_1^p(x,Q^2)$, including the most
recent E155 proton data \cite{ref24}.  The main reason for our LO
results being larger than the NLO ones in the small-$x$ region is 
due to the vanishing of $R^p(x,Q^2)$ in LO in (1.1).  The corresponding
$x$--dependence of $g_1^N(x,Q^2=5$ GeV$^2$) is shown in NLO in
Fig.\ 3 where the expected extrapolations into the yet unmeasured
small--$x$ region down to $x=10^{-3}$ are shown as well.  The solid
curves refer to our optimal NLO fit (with the input given in Table I)
and allowing our optimal total $\chi^2$ in Table I to vary by one
unit, $\delta \chi^2=\pm 1$, gives rise to the shaded areas due to
different choices of the polarized gluon input at $Q^2=\mu_{\rm NLO}^2$
in (2.6) such as $\delta g=\pm xg$, etc. In particular a vanishing
polarized gluon input $\delta g(x,\mu_{\rm NLO}^2)=0$, is for the 
time being entirely compatible with all present data as shown by the
dashed curves.  On the other hand, fully saturated (via the positivity
constraint) gluon inputs $\delta g(x,\mu_{\rm NLO}^2)=\pm g$ appear to be
disfavored by present data as shown by the dotted curves in Fig.\ 3.
It should be furthermore noted that the shaded bands in Fig.\ 3 contain
polarized gluon densities which correspond to first moments 
$\Delta g(Q^2=5$ GeV$^2)$ between $-0.81$ and $1.73$, according to input
moments between $\Delta g(\mu_{\rm NLO}^2)= -0.45$ and $0.7$, 
respectively, i.e.\ even negative total gluon polarizations are 
compatible with present data. The same results hold of course also in LO.
Future dedicated polarized small--$x$ measurements and upcoming
determinations of $\delta g(x,Q^2)$ should be useful
in removing such extrapolation ambiguities caused by our present
poor knowledge of $\delta g(x,Q^2)$. 

Our corresponding LO and NLO parton distributions at the
respective input scales $Q^2=\mu^2_{\rm LO,\, NLO}$ in Eq.\ (2.6)
with the `standard' scenario fit parameters given in \mbox{Table I} 
are shown in Fig.\ 4.  The main differences between our new input
densities and our old GRSV95 ones \cite{ref1} are somewhat harder
$\delta d$ (due to the new neutron data) and $\delta g$ 
distributions although, as discussed
above, the polarized gluon distribution in Fig.\ 4 is only
slightly preferred by our `optimal' fit to presently available
data.  The polarized input densities in Figs.\ 4(a) and 4(b) are
compared with our reference unpolarized valence--like LO and NLO
dynamical input densities of \cite{ref25}  which satisfy of
course the positivity requirement $|\delta f|\leq f$ as is
obvious from Eq.\ (2.6).  It should be nevertheless emphasized 
that the parameters, resulting from our rather general LO and
NLO fits, are always such that these positivity conditions are 
automatically satisfied, i.e.\ there is practically no need to
impose them separately.  The distributions at $Q^2=5$ GeV$^2$,
as obtained from these LO and NLO inputs at $Q^2=\mu^2$, are
shown in Fig.\ 5 where they are also compared with our old NLO
GRSV95 \cite{ref1} results.  It should be noted that the
substantially harder input polarized gluon density $x\delta
g(x,\mu^2)$, in particular in LO, in Fig.\ 4(a) as compared to our
old GRSV95 fit, causes the sea density $x\delta\bar{q}(x,Q^2)$ to
oscillate (slightly) in the large-$x$ region at $Q^2>\mu^2$ as
shown in Fig.\ 5 \cite{ref38}. 

Next let us turn to the first moments (total polarizations)
$\Delta f(Q^2)$ of our polarized parton distributions, as defined
in (1.3), and the resulting $\Gamma_1^{p,n}(Q^2)$ in (1.5).  It
should be recalled that, in contrast to the LO, the first moments
of the NLO (anti)quark densities do renormalize in the 
$\overline{\rm MS}$ scheme, i.e. are $Q^2$
dependent (see, e.g., Ref. \cite{ref1}), whereas the gluon
polarization $\Delta g(Q^2)$ renormalizes in both cases.  Our LO
`standard' fit implies
%Eq.(2.7)
\begin{eqnarray}
\Delta u = 0.871, & \quad & \Delta d = -0.396,
      \quad\quad \Delta\bar{q} = \Delta s = \Delta\bar{s} = -0.054\nonumber\\
\Delta g(\mu^2_{\rm LO}) = 0.190, & \quad & \Delta g (5\, {\rm GeV}^2) = 0.684,
      \quad\quad \Delta g(10\, {\rm GeV}^2) = 0.802
\end{eqnarray}
which result in $\Delta\Sigma = 0.259$ and
%Eq.(2.8)
\begin{equation}
\Gamma_1^p=0.151, \quad\quad \Gamma_1^n = -0.061\, .
\end{equation}
Our NLO results are summarized in Table II at some typical values of
$Q^2$.  Both our LO and NLO results for $\Gamma_1^{p,n}(Q^2)$ are
in satisfactory agreement with recent experimental determinations
[17 -- 24]. 
%\cite{ref17}--\cite{ref24}.  
Furthermore, due to the constraint
(2.1), the Bjorken sum rule \cite{ref39} holds manifestly in LO
and, according to (1.5), the NLO $\alpha_s$--corrected sum rule
reads
%Eq.(2.9)
\begin{equation}
\Gamma_1^p(Q^2)-\Gamma_1^n(Q^2) = \frac{1}{6}\, g_A
 \left( 1 - \frac{\alpha_s(Q^2)}{\pi} \right)\, .
 \end{equation}

It is also interesting to observe that at our low input scales
$Q^2=\mu^2_{\rm LO,\, NLO}=0.26,\, 0.40$ GeV$^2$ the nucleon's
spin carried by the total helicities of quarks and gluons amounts
only to \vspace{-0.5cm}
%Eq.(2.10)
\begin{eqnarray}
\frac{1}{2}\,\, \Delta\Sigma + \Delta g(\mu^2_{\rm LO}) & \simeq &
0.32\nonumber\\
\frac{1}{2}\,\, \Delta\Sigma_(\mu^2_{\rm NLO}) + \Delta
g(\mu^2_{\rm NLO}) & \simeq & 0.35
\end{eqnarray}
which implies for the helicity sum rule (1.9) already a sizeable
orbital contribution $L_{q+g}(\mu^2_{\rm LO,\, NLO})\simeq
0.18,\, 0.15$ at the low input scales.  Although this is in
contrast to our somewhat more intuitive previous GRSV95 result
\cite{ref1}, $L_{q+g}(\mu^2_{\rm LO,\, NLO})\simeq 0$, it should
be kept in mind that, for the time being, $\Delta g(Q^2)$ is
rather weakly constrained by present data as was discussed above.

Finally, for completeness we have also performed a NLO analysis in
a different factorization scheme, the so--called chirally invariant
(CI) or JET scheme \cite{ref8,ref40,ref41}, but any other choice
would do as well for studying the scheme dependence of our
$\overline{\rm MS}$ fit results.  Here, among other things 
\cite{ref8,ref40,ref41}, the total helicity of quarks, 
$\Delta\Sigma_{\rm CI}$,
is conserved, i.e.\ $Q^2$--independent, and is related to our 
$\Delta\Sigma$
in the $\overline{\rm MS}$ scheme via 
%Eq.(2.11)
\begin{equation}
\Delta\Sigma(Q^2) = \Delta\Sigma_{\rm CI} -3\, \frac{\alpha_s(Q^2)}{2\pi}\,
  \Delta g(Q^2)_{\rm CI}\, .
\end{equation}
Similarly agreeable fits as the ones in Figs.\ 1 -- 3 can be obtained
in this scheme, e.g.\, by choosing a large positive gluon density
with a total (input) helicity $\Delta g(\mu_{\rm NLO}^2)_{\rm CI} \simeq
0.6\, -\, 0.7$ being about three times larger than in Table II and the
sea density $\delta\bar{q}_{\rm CI}(\Delta\bar{q}_{\rm CI})$ turns out to be 
roughly 50\% smaller than the one of our best fit in the $\overline{\rm MS}$ 
`standard' scenario; here the total quark helicity increases to 
$\Delta\Sigma_{\rm CI}\simeq 0.4$.

\setcounter{equation}{0}
\section{SU(3)$_f$ broken `valence' (broken--sea) scenario}
The assumption of the flavor symmetric `standard' scenario with
its unbroken sea density in (2.5) is expected to be unrealistic, 
following our experience in the unpolarized case where a suppression 
of the strange sea component is required, as accomplished by the 
vanishing input $s(x,\mu^2) = \bar{s}(x,\mu^2) = 0$
in GRV98 \cite{ref25}, in order to comply with experimental
indications \cite{ref42,ref43} of an SU(3)$_f$ broken sea, and the
positivity constraint $\delta s\leq s$. Thus, in GRSV95 \cite{ref1}
we also considered a `valence' scenario where, in contrast to (2.5),
%Eq.(3.1)
\begin{eqnarray}
\delta\bar{q}(x,\mu^2) & \equiv & \delta\bar{u} = \delta\bar{d} =
  \delta u_{\rm{sea}} = \delta d_{\rm{sea}}\nonumber\\
 \delta s(x,\mu^2) &  = & \delta\bar{s}(x,\mu^2)  = 0\, .
\end{eqnarray}
Furthermore the full SU(3)$_f$ flavor symmetry, giving rise to
the constraints (2.1) and (2.2), is broken in the `valence'
scenario to the extent \cite{ref44} that the flavor--changing
hyperon $\beta$--decay data fix {\em{only}} the total
helicity of {\em{valence}} quarks $\Delta q_v \equiv \Delta
q -\Delta\bar{q}$:
%Eqs.(3.2)+(3.3)
\begin{eqnarray}
\Delta u_v(\mu^2) - \Delta d_v(\mu^2) & = & F+D\\
\Delta u_v(\mu^2) + \Delta d_v(\mu^2) & = & 3F-D\, ,
\end{eqnarray}
i.e.\ $\Delta q_3 = \Delta u_v - \Delta d_v$ and $\Delta q_8 =
3F-D + 4\Delta\bar{q}$ at $Q^2=\mu^2$ \cite{ref1} according to
(3.1).  Therefore a light polarized sea $\Delta\bar{q}<0$
suffices here to account for the reduction of the Ellis--Jaffe
estimate (2.4).  This is the reason for our simplifying
assumption of a maximally broken SU(3)$_f$ strange sea input in
(3.1) in order to reduce the number of input distributions to be
fitted to the rather scarce available polarization data, which
are now sufficient for fixing these input distributions.
(Future high--statistics data should allow, at least in principle,
to extract the total strange sea polarization without employing
any simplifying assumption, as for example from (1.8),
$\Delta(s+\bar{s})=(\Delta\Sigma -\Delta q_8)/3\,$.) The
quality of the fits obtained is comparable to that of the
`standard' flavor--symmetric scenario discussed and presented in
the previous Section, cf.\ Fig.\ 1.  We refrain, however, from
presenting these results here explicitly because the assumed
remaining flavor--isospin symmetry of the light sea components in
(3.1) appears to be somewhat artificial and unnatural in view of
the flavor--asymmetric unpolarized light sea distributions
$\bar{d}(x,Q^2)>\bar{u}(x,Q^2)$ [25 -- 28]. 
%\cite{ref25}--\cite{ref28}.

Turning now to the presumably more realistic scenario where also the
{\em{flavor--isospin symmetry of the polarized sea is broken}}, we note, 
as already pointed out in the Introduction, that some model assumptions
are needed for the corresponding input distributions.  The analysis
of the unpolarized structure functions yields
%Eq.(3.4)
\begin{equation}
\bar{d}(x,\mu^2)/ \bar{u}(x,\mu^2)\simeq u(x,\mu^2)/d(x,\mu^2),
\end{equation}
which holds to a rather good accuracy for the GRV98 distributions
\cite{ref25}.  This proportionality relation is expected to hold
approximately at least for  $0.01 \lesssim x \lesssim 0.3$
where the breaking of
the light sea $\bar{d}>\bar{u}$ is directly tested experimentally
via Drell--Yan dilepton production in $pp$ and $pd$ collisions
\cite{ref45} and semi--inclusive $\pi^{\pm}$ production in $ep$
and $ed$ reactions \cite{ref46}. The relation (3.4) may be
considered \cite{ref30} as a manifestation of the Pauli--blocking
effect \cite{ref47} which should be relevant also in the
polarized parton sector.  We therefore estimate the
flavor--symmetry breaking of the polarized sea to be given in a 
first approximation by
\cite{ref30}
%Eq.(3.5)
\begin{equation}
\delta\bar{d}(x,\mu^2)/\delta \bar{u}(x,\mu^2) = \delta u(x,\mu^2)/
    \delta d(x,\mu^2)
\end{equation}
together with the previously advocated
%Eq.(3.6)
\begin{equation}
\delta s(x,\mu^2) = \delta\bar{s}(x,\mu^2)=0\, .
\end{equation}
According to the unpolarized case above, we expect the
proportionality relation (3.5) to hold approximately at least for
$0.01 \lesssim x \lesssim 0.3$.

It should be reemphasized that, in complete analogy to unpolarized
DIS, data on inclusive polarized DIS in kinematical regimes where
only photon exchange is relevant give information only on the
sums of quark and antiquark polarizations for each flavor, $\delta
q(x,Q^2)+\delta\bar{q}(x,Q^2)$, appearing in $g_1(x,Q^2)$ in
(1.2). This implies, in particular, that the amount of
flavor--SU(2) breaking in the polarized sea cannot be determined
from such data since one can always change the parton densities
by $\delta q\to\delta q - \delta\phi_q$ and
$\delta\bar{q}\to\delta\bar{q} +\delta\phi_q$ for any arbitrary
functions $\delta\phi_u(x)$ and $\delta\phi_d(x)$
{\em{without}} affecting at all the measured structure
functions $g_1^{p,n,d}(x,Q^2)$.  Restrictions on $\delta\phi_q(x)$
occur only in our `valence' scenario where $\Delta\phi_d = -\Delta\phi_u$ 
for the first moments, as in general implied by our modified
$(\Delta\bar{u}\neq\Delta\bar{d})$ valence scenario constraints 
(3.7) and (3.8) below, and which have to even vanish for the
`unbroken' $(\Delta \bar{u}=\Delta\bar{d})$ constraints (3.2) and
(3.3). 
Therefore, for the time being, one has to resort to some, as far
as possible, general and not too restrictive model assumptions
concerning the breaking of the flavor--symmetry of the polarized
light sea ($\delta\bar{u}\neq\delta\bar{d}$), such as the
proportionality (3.5).  Only future polarized $\vec{p}\,\vec{p}$
and $\vec{p}\,\vec{d}$ Drell--Yan $\mu^+\mu^-$ pair and weak vector
boson production experiments \cite{ref48}  as well as polarized 
semi--inclusive
DIS $\vec{e}\,\vec{p}\,(\vec{d}\,)\to e\,(\pi,K)X$ experiments 
\cite{ref49,ref50}
can provide us with direct measurements of the individual
$\delta\bar{u}(x,Q^2)$ and $\delta\bar{d}(x,Q^2)$ distributions.

We now have, instead of (3.2),
%Eq.(3.7)
\begin{equation}
\Delta q_3  =  \Delta u_v(\mu^2)-\Delta d_v(\mu^2) +
2[\Delta\bar{u}
   (\mu^2)-\Delta\bar{d}(\mu^2)] = F+D\, ,
\end{equation}
%
%Eq.(3.8)
 and on account of (3.3) and (3.6)
\begin{eqnarray}
 \Delta q_8(\mu^2) & = & \Delta u_v(\mu^2) + \Delta d_v(\mu^2) +
     2[ \Delta\bar{u}(\mu^2)+\Delta\bar{d}(\mu^2)]\nonumber\\
 & = & 3F-D + 2[\Delta\bar{u}(\mu^2)+\Delta\bar{d}(\mu^2)]\nonumber\\
 & = & \Delta\Sigma(\mu^2)
\end{eqnarray}
where the imposed constraint (3.7) guarantees that the Bjorken
sum rule (2.9) holds manifestly.  Thus Eq.\ (1.5) becomes
%Eq.(3.9)
\begin{equation}
\Gamma_1^{p,n}(Q^2) = \left[ \pm \frac{1}{12} (F+D) + \frac{5}{36} (3F-D)
  + \frac{10}{36} \left(\Delta\bar{u}(Q^2)+\Delta\bar{d}(Q^2)\right)
     \right] \left( 1 - \frac{\alpha_s(Q^2)}{\pi}\right)
\end{equation}
apart from a marginal contribution $\Delta
s(Q^2)=\Delta\bar{s}(Q^2)<0$ which is generated dynamically via
the NLO evolution to $Q^2>\mu_{\rm{NLO}}^2$ even for the
vanishing input in (3.6).  Thus, in this case, only the
{\em{total}} light--quark sea contribution in (3.9) has to be
{\em{negative}}, $\Delta\bar{u}(Q^2) +\Delta\bar{d}(Q^2)<0$, 
in order to achieve the experimentally required reduction of the 
Ellis--Jaffe expectation (2.4).

Our resulting input distributions can be parametrized as in (2.6)
where now, instead of the unbroken $\delta\bar{q}$ sea, we have a 
similar parametrization for $\delta\bar{u}(x,\mu^2)$ and $\delta\bar{d}
(x,\mu^2)$ which are constrained by (3.5) \cite{ref51}, together with
the respective flavor--broken unpolarized input densities $\bar{u}(x,\mu^2)$
und $\bar{d}(x,\mu^2)$ taken from \cite{ref25}.  The parameters of our
optimal LO densities at $\mu_{\rm LO}^2 = 0.26$ GeV$^2$ and the ones
of the NLO($\overline{\rm MS}$) densities at $\mu_{\rm NLO}^2= 0.40$
GeV$^2$ are given in Table III.  These optimal fits correspond to 
$\chi_{\rm DF,\, LO}^2=0.823$ and $\chi_{\rm DF,\, NLO}^2 = 0.816$
similarly to the  `standard' scenario in Sec.\ 2.  The quality of these
LO and NLO fits to $A_1^N(x,Q^2)$ in the broken  `valence' scenario
is practically identical to our new fit in the `standard' scenario
shown in Fig.ž 1 by the solid curves. The same holds true also for 
$g_1^{p,n,d}(x,Q^2)$ shown in Figs.\ 2 und 3.  The corresponding LO
and NLO parton distributions at the respective input scales  
$Q^2=\mu_{\rm LO,\, NLO}^2$ in Eq.\ (2.6) \cite{ref51} with the 
`valence' scenario fit parameters given in Table III are shown in
Fig.\ 6, which are also compared with our reference unpolarized
valence--like dynamical input densities of \cite{ref25} which satisfy
the positivity constraint $|\delta f|\leq f$. The polarized gluon
densities turn out to be somewhat larger here, in particular in NLO, 
than
the ones in the `standard' scenario shown in \mbox{Fig.\ 4}.  It should
be
furthermore emphasized that we always expect for the broken light--sea
input densities to have a {\em{positive}} $\delta{\bar{u}}$ and a 
{\em{negative}} $\delta\bar{d}$ with 
$|\delta\bar{d}|>\delta\bar{u}$, i.e.\ $\delta\bar{u}-\delta\bar{d}>0$
and $\delta\bar{u} +\delta\bar{d}<0$.

In Fig.\ 7 we present the flavor asymmetry $x(\delta\bar{u}-\delta\bar{d})
(x,\mu^2)$ separately, as obtained from Fig.\ 6, which compares
favorably with predictions of the relativistic field theoretical
chiral quark--soliton model \cite{ref52,ref50}.  Similar results 
have been obtained by a recent analysis \cite{ref53} based on the 
statistical parton model which are supposed to hold at a somewhat
larger input scale $Q_0^2={\rm M}_p^2\simeq 0.9$ GeV$^2$.  We note
furthermore that the prediction $\delta g(x,{\rm M}_p^2) = 0$ of
this latter model is consistent with the results of the present
analysis which do {\em{not}} exclude this possibility, cf.\
Fig.\ 3.  The resulting NLO distributions at $Q^2=5$ GeV$^2$ are 
shown in Fig.\ 8 where they are also compared with our new NLO
results obtained in the `standard' scenario as shown by the solid
curves in Fig.\ 5.   

The first moments (total polarizations) $\Delta f(Q^2)$ of the 
polarized parton distributions of our fully flavor--broken `valence'
scenario and the resulting $\Gamma_1^{p,n}(Q^2)$ are in LO given by 
%Eq.(3.10)
\begin{equation}
\renewcommand{\arraystretch}{1.5}
  \begin{array}[b]{c}
\Delta u = 0.664, \quad \Delta d = -0.248, \quad
        \Delta\bar{u} = 0.093, \quad  \Delta\bar{d} = -0.261, \quad
       \Delta s = \Delta\bar{s} = 0\\ 
\Delta g(\mu^2_{\rm LO}) = 0.300,  \quad  \Delta g (5\, {\rm GeV}^2) = 0.963,
      \quad   \Delta g(10\, {\rm GeV}^2) = 1.122
  \end{array}
\end{equation}
which result in $\Delta\Sigma = 0.248$ and
%Eq.(3.11)
\begin{equation}
\Gamma_1^p=0.140, \quad\quad \Gamma_1^n = -0.071\, .
\end{equation}
Our NLO results are summarized in Table IV at some typical values of $Q^2$.
The nucleon's spin is carried almost entirely by the total helicities
of quarks and gluons at the LO and NLO input scales
%Eq.(3.12)
\begin{equation}
\frac{1}{2}\Delta\Sigma(\mu_{\rm LO,\, NLO}^2) +\Delta g(\mu_{\rm LO,\, NLO}^2)
  \simeq 0.42, \,\,\, 0.48
\end{equation}
which is larger than the  `standard' scenario results (2.10), and thus a 
very small orbital contribution $L_{q+g}(\mu_{\rm LO,\, NLO}^2)\simeq 0.08$, 0.02
is required at the low input scales in order to comply with the sum rule (1.9).
This is somewhat similar to our previous results \cite{ref1}, but again
$\Delta g(Q^2)$ is not strongly constrained by present data.  Nevertheless
it is intuitively appealing that this nonperturbative orbital (angular
momentum) contribution to the helicity sum rule (1.9) vanishes at our low
input scales, $L_{q+g}(\mu^2)\simeq 0$.  This is in contrast to larger scales
$Q^2>\mu^2$ where hard radiative effects give rise to sizeable orbital
components due to the increasing $k_T$ of the partons, which eventually
have to compensate in (1.9) the strongly increasing gluon polarization 
$\Delta g(Q^2)\sim \alpha_{s}^{-1}(Q^2)$ : in both scenarios we obtain,
for example, $\Delta g(10\,\, {\rm GeV}^2)\simeq 1$.

Finally let us conclude with a few remarks concerning the flavor--symmetry
breaking which was implemented in our broken `valence' scenario via
the entirely empirical relation (3.5).  On rather general grounds one
expects the product
%Eq.(3.13)
\begin{equation}
\delta q(x,\mu^2)\, \delta\bar{q}(x,\mu^2) \equiv P(x)
\end{equation}
to be a universal flavor--{\em{in}}dependent function $P(x)$,
since the effect of Pauli--blocking is only related to the spin
(helicities) of quarks and antiquarks irrespective of their flavor
degree of freedom.  This implies $\delta u(x,\mu^2)\,\delta\bar{u}
(x,\mu^2)=\delta d(x,\mu^2)\,\delta\bar{d}(x,\mu^2)$, i.e.\ Eq.\ (3.5).
Furthermore, we have seen that the data select, within our `valence'
scenario with its totally flavor--broken polarized light sea densities
in (3.5) and (3.6), the solution of 
Eq.\ (3.5) which satisfies $P(x)>0$ in (3.13) for $q=u,\, d$ as can
be seen in Fig.\ 6.  This can be understood \cite{ref30} as a 
consequence of the expected predominant {\em{pseudoscalar}}
configuration \cite{ref29,ref54} of the quark--antiquark pairs in
the nucleon sea.  In fact, the two relations $u\bar{u}\simeq d\bar{d}$
and $\delta u\,\delta\bar{u}=\delta d\, \delta\bar{d}$ at the input scale
$Q^2=\mu^2$ can be rewritten as
%Eq.(3.14)  
\begin{eqnarray}
P_p(x) & \equiv & u_+\bar{u}_+ + u_-\bar{u}_-\,
          \simeq\,  d_+\bar{d}_+ + d_-\bar{d}_-\nonumber\\
P_a(x) & \equiv & u_+\bar{u}_- + u_-\bar{u}_+\, 
          \simeq\,  d_+\bar{d}_- + d_-\bar{d}_+
\end{eqnarray}
with $P=P_p -P_a$ in (3.13) and the common helicity densities being
given by 
$\stackrel{(-)}{q}_{\pm}=(\stackrel{(-)}{q}\!\!\pm\,\, \delta\!\!\!
\stackrel{(-)}{q})/2$ where for brevity we have dropped the 
$x$--dependence.  A predominant pseudoscalar configuration of 
$(q\bar{q})$--pairs in the nucleon sea implies, via Pauli--blocking,
that the aligned quark--quark configurations $q_+(q_+\bar{q}_-)$ and
$q_-(q_-\bar{q}_+)$ are suppressed relatively to the antialigned
$q_+(q_-\bar{q}_+)$ and $q_-(q_+\bar{q}_-)$ `cloud' configurations,
i.e.\ $P_p(x)>P_a(x)$ which implies $P(x)>0$ in (3.13).  The result
for $P_p/P_a$, corresponding to our optimal fit, is shown in Fig.\ 9:
clearly, this ratio will be maximal where $xq(x,\mu^2)$ and
$x\delta q(x,\mu^2)$ are maximal at $x\simeq 0.2\,-\,0.4$, cf.\
Fig.\ 6, i.e.\ where the Pauli--blocking, Eq.\ (3.13), is most
effective which is nicely exhibited in Fig.\ 9 in LO and NLO.

It is interesting to mention that some of these expectations, which
derive mainly from our light--sea flavor breaking relation (3.5),
have been already confirmed by a recent entirely independent
simultaneous analysis \cite{ref14} of polarized DIS and semi--inclusive
deep inelastic scattering (SIDIS) asymmetry--data.  In particular 
the more recent high precision SIDIS HERA--HERMES data \cite{ref55} on
$h^+$ production ($h=\pi,$K dominantly) off a proton target,
$\vec{e}\,\vec{p}\to eh^+X$, seem to play a decisive role in
favoring flavor--broken light sea densities $\delta\bar{u}(x,Q^2)\neq
\delta\bar{d}(x,Q^2)$, despite the fact that these asymmetry data
on $A_{1p}^{h^+}$ refer to rather small scales 
$Q^2$ \raisebox{-0.1cm}{$\stackrel{>}{\sim}$} 1 GeV$^2$. 
The reason for this discriminative power is, when combined with the 
data from inclusive DIS, due to the fact that
$A_{1p}^{h^+}$ is proportional \cite{ref14}, besides to the dominant
valence  contribution, also to $\delta\bar{d} -4\delta\bar{u}$,
multiplied by a `favored' fragmentation function, which is significantly
more sensitive to $\delta\bar{u}$ than to $\delta\bar{d}$.  A clear
preference for a {\em{positive}} $\delta\bar{u}$ has been
observed \cite{ref14}, which is very similar to our NLO $\delta\bar{u}$
shown in Fig.\ 6(b), and a flavor symmetric  `standard' light sea
scenario seems to be strongly disfavored.

We have calculated at NLO the spin asymmetries for semi--inclusive DIS
using the well known theoretical SIDIS framework  
\cite{ref56,ref14} together with our results for 
the polarized parton distributions of the `standard' and `valence'
scenario with their flavor--symmetric and flavor--broken light sea
densities, respectively, employing the fragmentation functions 
of \cite{ref57}. (We did not use the alternative set of recent
fragmentation functions suggested in  \cite{ref58}, since they 
refer to scales $Q^2$ larger than 2 GeV$^2$.)  The results for the
relevant SIDIS asymmetry $A_{1p}^{h^+}$ are shown in Fig.\ 10.  
Although the high precision HERMES data \cite{ref55} seem to favor 
slightly the `valence' scenario with its flavor--broken light sea,
the results of the `standard' scenario with its flavor--symmetric
light sea cannot yet be ruled out. Both scenarios in Fig.\ 10 give
rise to a comparable $\chi^2/$(9 data points) of  7.6 and 8.5 for
the `valence' and `standard' scenario, respectively.

%Sec. 4, Summary and conclusions
\section{Summary and conclusions}
All recent polarized DIS data, including the most recent SLAC-E155
proton data \cite{ref24}, have been analyzed and studied within 
the `standard' and `valence' scenario in LO and NLO of QCD.  The 
`standard' scenario, characterized by (2.1) and (2.2),  refers to 
the common simplified, but probably 
unrealistic, assumption of an SU(3)$_f$ flavor--symmetric polarized 
light sea. The original `valence' scenario \cite{ref1}, characterized 
by (3.2) and (3.3), is now modified by employing a {\em{totally}}  
SU(3)$_f$ asymmetric polarized light sea 
$\delta\bar{u}\neq\delta\bar{d}\neq\delta\bar{s}$ which leads to the 
modified constraints (3.7) and (3.8).  Since inclusive
 polarized DIS data cannot fix the flavor--broken sea
densities, we have modelled the flavor--asymmetric light sea
densities $\delta\bar{u}\neq\delta\bar{d}$ using a Pauli--blocking
ansatz \cite{ref30} in (3.5), because similar  `blocking' effects 
can
also explain the flavor--asymmetry of unpolarized sea densities 
($\bar{d}>\bar{u}$).  All our resulting polarized parton distributions
respect the fundamental positivity constraints down to the low 
resolution scales $Q^2=\mu_{\rm LO}^2=0.26$ GeV$^2$ and $\mu_{\rm NLO}^2
=0.40$ GeV$^2$.  The polarized gluon distribution $\delta g(x,Q^2)$ is
weakly constrained by present data in both scenarios.  In particular,
a vanishing gluon input $\delta g(x,\mu^2)=0$ is equally compatible
with all present measurements of $A_1^{N}(x,Q^2)$ or $g_1^{N}
(x,Q^2)$. Only a fully saturated (via the positivity constraint)
gluon input $\delta g(x,\mu^2)=\pm g(x,\mu^2)$ appears to be disfavored 
by present data.

The presumably more realistic `valence' scenario with its flavor--broken
light sea quark distributions $\delta \bar{u}\neq\delta\bar{d}\,
(\neq \delta\bar{s}=\delta s\simeq 0)$ leads to a {\em{positive}}
$\delta\bar{u}(x,Q^2)$ density and a sizeably larger negative
$\delta\bar{d}(x,Q^2)$.  These results are supported by a recent
combined analysis \cite{ref14} of polarized DIS and  semi--inclusive
DIS data and agree with predictions of the relativistic field 
theoretical chiral quark--soliton model \cite{ref52,ref50} and of the 
statistical parton model \cite{ref53}.  Present high statistics HERA--HERMES
data \cite{ref55} on semi--inclusive asymmetries $A_{1{\rm N}}^{h^{\pm}}$
for $h^{\pm}$ production off nucleon targets cannot, however, yet
uniquely discriminate between our  `valence' scenario with flavor--broken
polarized light sea densities and the common `standard' scenario with
a flavor--symmetric light sea--quark distribution.

A FORTRAN package containing our optimally fitted  `standard' and
fully flavor--broken `valence' NLO($\overline{\rm MS}$) as well as
LO distributions can be obtained by electronic mail.
\vspace{1.0cm}        

\noindent{\large\bf{Acknowledgements}}

We are grateful to Greg Mitchell for useful information concerning
the recent E155 proton data.
This work has been supported in part by the `Bundesministerium
f\"ur Bildung, Wissenschaft, Forschung und Technologie'
and the `Deutsche Forschungsgemeinschaft', Berlin/Bonn.
W.V.\ is grateful to RIKEN, Brookhaven National
Laboratory and the U.S.\ Department of Energy (contract number
DE-AC02-98CH10886) for providing the facilities essential for the 
completion of his work.

\newpage

%%%%%%%
%TABLES
%%%%%%%
%
%%%%%%%%%
% TABLE I
%%%%%%%%%
\noindent{\large{\bf{\underline{Table I.}}}}\hspace{0.25cm} The
parameters of the LO and NLO input parton distributions, as defined
in Eq.\ (2.6), at $\mu_{\rm LO}^2=0.26$ GeV$^2$ and $\mu_{\rm NLO}^2
=0.40$ GeV$^2$, respectively, for the `standard' scenario as obtained
from fits to the data in [15 -- 24].
\begin{table}[ht]
\begin{center}
\vspace*{-0.3cm}
\begin{tabular}{|c||c|c|}
\hline
& \multicolumn{2}{c|}{standard scenario}
\\ \cline{2-3}
& LO & NLO \\ \hline\hline
$N_u$      & 0.851  & 1.019 \\
$\alpha_u$ & 0.45      & 0.52    \\ 
$\beta_u$  & 0         & 0.12    \\ \hline
$N_d$      & -0.734  & -0.669 \\
$\alpha_d$ & 0.49      & 0.43    \\ 
$\beta_d$  & 0.03      & 0       \\ \hline
$N_{\bar{q}}$ & -0.587 & -0.272 \\
$\alpha_{\bar{q}}$ & 0.68  &  0.38    \\
$\beta_{\bar{q}}$  & 0     &  0       \\ \hline
$N_s$ & 1 & 1 \\ \hline
$N_g$      & 1.669     & 1.419 \\
$\alpha_g$ & 1.79       & 1.43 \\
$\beta_g$  & 0.15      & 0.15  \\ \hline \hline
$\chi^2/209$ data pts. & 174.9 & 169.8 \\ \hline
\end{tabular}
\end{center}
\end{table}
\vspace*{-0.5cm}

%%%%%%%%%%
% TABLE II
%%%%%%%%%%
%
\noindent{\large{\bf{\underline{Table II.}}}}\hspace{0.25cm} 
First moments (total polarizations) $\Delta f$ of polarized NLO
parton densities $\delta f(x,Q^2)$ and $g_1^{p,n}(x,Q^2)$, defined
in (1.3) and (1.4), as obtained in the `standard' scenario.  The
marginal differences between $\Delta\bar{u}$ and $\Delta\bar{d}$
at $Q^2>\mu^2$, generated dynamically by the NLO evolution, are
not displayed.
\begin{table}[ht]
\begin{center}
\vspace*{-0.5cm}
\renewcommand{\arraystretch}{1.2}
\begin{tabular}{|c||c|c|c|c|c||c|c|}
\hline
$Q^2\,({\rm{GeV}}^2)$ & $\Delta u$ & $\Delta d$ & $\Delta \bar{q}$ &
$\Delta g$ & $\Delta \Sigma $&
$\Gamma_1^p$ & $\Gamma_1^n$ \\ \hline
$\mu^2_{NLO}$ & 0.863 & -0.404 & -0.062 & 0.240 & 0.211 & 0.119 & -0.054\\
1             & 0.861 & -0.405 & -0.063 & 0.420 & 0.204 & 0.127 & -0.058\\
5             & 0.859 & -0.406 & -0.064 & 0.708 & 0.197 & 0.132 & -0.062\\
10            & 0.859 & -0.406 & -0.064 & 0.828 & 0.197 & 0.133 & -0.063\\
\hline
\end{tabular}
\end{center}
\end{table}
\newpage

%
%%%%%%%%%%%
% TABLE III
%%%%%%%%%%%
%
\noindent{\large{\bf{\underline{Table III.}}}}\hspace{0.25cm} The
parameters of the LO and NLO input distributions, as defined
in Eq.\ (2.6), with $\bar{q}$ to be identified with $\bar{u}$,  
at $\mu_{\rm LO}^2=0.26$ GeV$^2$ and $\mu_{\rm NLO}^2=0.40$ GeV$^2$, 
respectively, for the `valence' scenario.   We have fitted the
broken light--sea input density $\delta\bar{u}(x,\mu^2)$ and fixed
$\delta\bar{d}(x,\mu^2)$ via (3.5) and (3.4) \cite{ref51}. 
\begin{table}[ht]
\begin{center}
\vspace*{-0.3cm}
\begin{tabular}{|c||c|c|}
\hline
& \multicolumn{2}{c|}{valence scenario}
\\ \cline{2-3}
& LO & NLO \\ \hline\hline
$N_u$      & 1.297     & 2.043  \\
$\alpha_u$ & 0.79      & 0.97   \\ 
$\beta_u$  & 0.27      & 0.64   \\ \hline
$N_d$      & -2.496    & -2.709 \\
$\alpha_d$ & 1.17      & 1.26   \\ 
$\beta_d$  & 1.31      & 1.06       \\ \hline
$N_{\bar{u}}$ & 2.005 & 1.727   \\
$\alpha_{\bar{u}}$ & 0.79  &  0.73    \\
$\beta_{\bar{u}}$  & 1.93  &  2.00     \\ \hline
$N_s$ & 0 & 0 \\ \hline
$N_g$      & 6.637     & 20.45 \\
$\alpha_g$ & 2.00      & 2.92  \\
$\beta_g$  & 1.50      & 1.68  \\ \hline \hline
$\chi^2/209$ data pts. & 172.0 & 170.5 \\ \hline
\end{tabular}
\end{center}
\end{table}
\vspace*{-0.5cm}

%%%%%%%%%%
% TABLE IV
%%%%%%%%%%
%
\noindent{\large{\bf{\underline{Table IV.}}}}\hspace{0.25cm} 
First moments (total polarizations) $\Delta f$ of polarized 
parton densities $\delta f(x,Q^2)$ and $g_1^{p,n}(x,Q^2)$, defined
in (1.3) and (1.4), as obtained in the fully flavor--broken 
`valence' scenario.  The marginal finite $\Delta s = \Delta\bar{s}
(Q^2>\mu_{\rm NLO}^2)$ are generated dynamically by the NLO 
evolution.
\begin{table}[ht]
\begin{center}
\vspace*{-0.5cm}
\renewcommand{\arraystretch}{1.2}
\begin{tabular}{|c||c|c|c|c|c|c|c||c|c|}
\hline
$Q^2\,({\rm{GeV}}^2)$ & $\Delta u$ & $\Delta d$ & $\Delta \bar{u}$ & $\Delta \bar{d}$ &
$\Delta s=\Delta \bar{s}$ & $\Delta g$ & $\Delta \Sigma $&
$\Gamma_1^p$ & $\Gamma_1^n$ \\\hline
$\mu^2_{NLO}$ & 0.693 & -0.255 & 0.087 & -0.232 & 0 & 0.330 & 0.293 & 0.119 & -0.053  \\
1             & 0.691 & -0.257 & 0.086 & -0.234 & $-1.95\times 10^{-3}$& 
0.579& 0.282 & 0.127 & -0.058 \\
5             & 0.689 & -0.258 & 0.085 & -0.235 & $-3.31\times 10^{-3}$& 
0.974& 0.273 & 0.132& -0.062 \\
10            & 0.688 & -0.258 & 0.085 & -0.236 & $-3.64\times 10^{-3}$& 
1.140& 0.272 & 0.133& -0.062  \\ \hline
\end{tabular}
\end{center}
\end{table}
\newpage

%%%%%%%%%%%%%%%%
%FIGURE CAPTIONS
%%%%%%%%%%%%%%%%
\noindent
{\large{\bf{\underline{Figure Captions}}}}

\begin{itemize}
\item[\bf{Fig.\ 1}.]  Comparison of our NLO results for $A_1^N(x,Q^2)$
      as obtained from the fitted input at $Q^2=\mu_{\rm NLO}^2$ for
      the  `standard' SU(3)$_f$ flavor--symmetric sea scenario [Eq.\
      (2.6) and Table I] with present data [16 -- 24].
      %\cite{ref16}--\cite{ref24}  
      The $Q^2$ values adopted here correspond to the different
      values quoted in [16 -- 24] %\cite{ref16}--\cite{ref24}
      for each data point starting at $Q^2\geq 1$ GeV$^2$ at the 
      lowest available $x$ bin.  Our old NLO GRSV95 fit \cite{ref1}
      is shown for comparison as well (dashed curves).  Our present
      LO fit is very similar to the NLO one shown by the solid 
      curves.

\item[\bf{Fig.\ 2}.]  Comparing the $Q^2$--dependence of our LO and
      NLO `standard' scenario results with recent data [16 -- 18, 21, 24]
      %\cite{ref16}--\ref{18}, \cite{ref21,ref24}
      on $g_1^p(x,Q^2)$.  To ease the graphical representation we
      have multiplied all results at the various fixed values of $x$
      by the numbers indicated in parentheses.

\item[\bf{Fig.\ 3}.]  The $x$--dependence of $g_1^N$ at $Q^2=5$ GeV$^2$
      in the NLO `standard' scenario.  Different choices of the gluon
      input $\delta g(x,\mu_{\rm NLO}^2)$ in (2.6) are shown by the
      dashed and dotted curves as indicated.  Allowing our optimal
      total $\chi^2$ to change by one unit, $\delta\chi^2=\pm 1$,
      results in the shaded areas shown.  The data are taken from
      [17, 18, 20, 22 -- 24].
      %\cite{ref17,ref18,ref20},\cite{ref22}--\cite{ref24}

\item[\bf{Fig.\ 4}.]  {\bf (a)} Comparison of our fitted  `standard' 
      LO input distributions in Eq.\ (2.6) and Table I at 
      $\mu_{\rm LO}^2=0.26$ GeV$^2$ with our previous old GRSV95 fit
      \cite{ref1} and with the unpolarized dynamical GRV98 input
      densities of  \cite{ref25}.  {\bf (b)} The same as in (a) but for the
      NLO input densities at $\mu_{\rm NLO}^2=0.40$ GeV$^2$.

\item[\bf{Fig.\ 5}.]  The polarized LO and NLO($\overline{\rm MS}$) 
      distributions at $Q^2=5$ GeV$^2$ in the `standard' scenario,
      as obtained from the input densities at 
      $Q^2=\mu_{\rm LO,\, NLO}^2$ in Fig.\ 4.  
      Our old NLO results \cite{ref1}
      are shown for comparison.

\item[\bf{Fig.\ 6}.] {\bf (a)} Comparison of our fitted fully broken `valence'
      scenario LO input distributions in Eq.\ (2.6) \cite{ref51}
      and Table III at $\mu_{\rm LO}^2=0.26$ GeV$^2$ with the
      unpolarized GRV98 input densities of \cite{ref25}.  {\bf (b)}  The
      same as in (a) but for the NLO input densities at 
      $\mu_{\rm NLO}^2 = 0.40$ GeV$^2$.

\item[\bf{Fig.\ 7}.] LO and NLO results for the difference of the
      broken light--sea input densities $\delta\bar{u}(x,\mu^2)$ and
      $\delta\bar{d}(x,\mu^2)$ in the  `valence' scenario as obtained
      from Fig.\ 6.  The prediction of the chiral quark--soliton 
      model is taken from K.\ Goeke et al. \cite{ref52}.

\item[\bf{Fig.\ 8}.] The polarized LO and NLO($\overline{\rm MS}$)
      distributions at $Q^2=5$ GeV$^2$ in the fully flavor--broken
      `valence' scenario, as obtained from the input densities at
      $Q^2=\mu_{\rm LO,\, NLO}^2$ in Fig.\ 6.  For comparison the
      new NLO results of the `standard' (unbroken sea) scenario are
      shown as well by the dashed curves (which coincide with the
      solid curves of Fig.\ 5.).

\item[\bf{Fig. 9}.]  The ratio of $P_p(x)$ and $P_a(x)$, defined in
      (3.14), at the LO and NLO input scales $\mu_{\rm LO,\, NLO}^2$
      which demonstrates the Pauli--blocking of the disfavored 
      antiparallel $q_{\pm}\bar{q}_{\mp}$ configurations relative
      to the favored parallel $q_{\pm}\bar{q}_{\pm}$ configurations
      as discussed in the text.

\item[\bf{Fig.\ 10}.] NLO predictions for the semi--inclusive DIS
      asymmetry $A_{1p}^{h^+}$ for $h^+$ production off a proton
      target within the `valence' and `standard' scenario.  The
      HERMES data are taken from \cite{ref55}.
\end{itemize}
%
%%%%%%%%%%%%%%%%%%%%%%%%%%%
%        FIGURES
%%%%%%%%%%%%%%%%%%%%%%%%%%%
%
\newpage
\pagestyle{empty}

\begin{center}
\epsfig{file=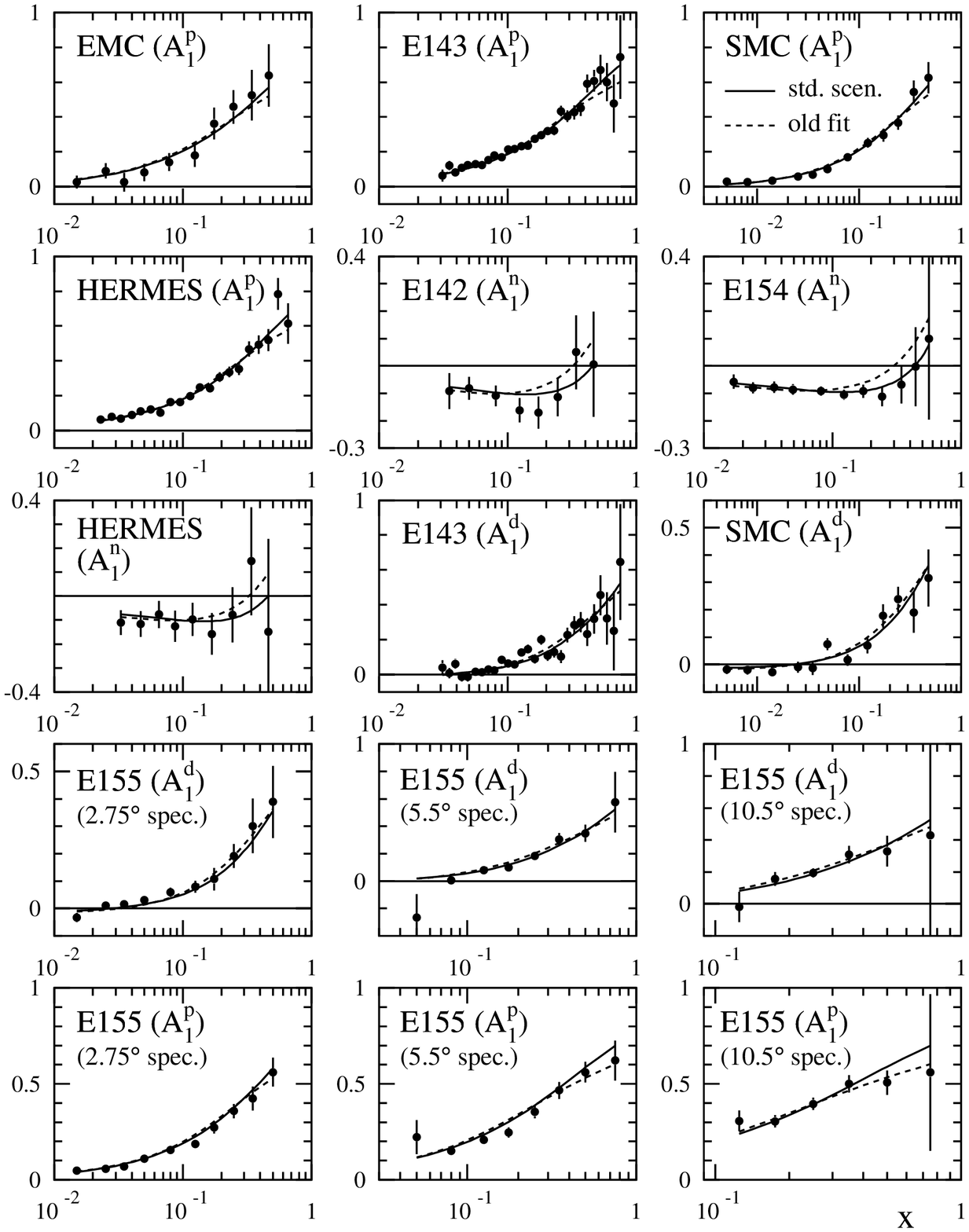,width=0.99\textwidth}
\large {\bf Fig.\ 1} \normalsize
\end{center}
\newpage

\vspace*{-1.5cm}
\begin{center}
\epsfig{file=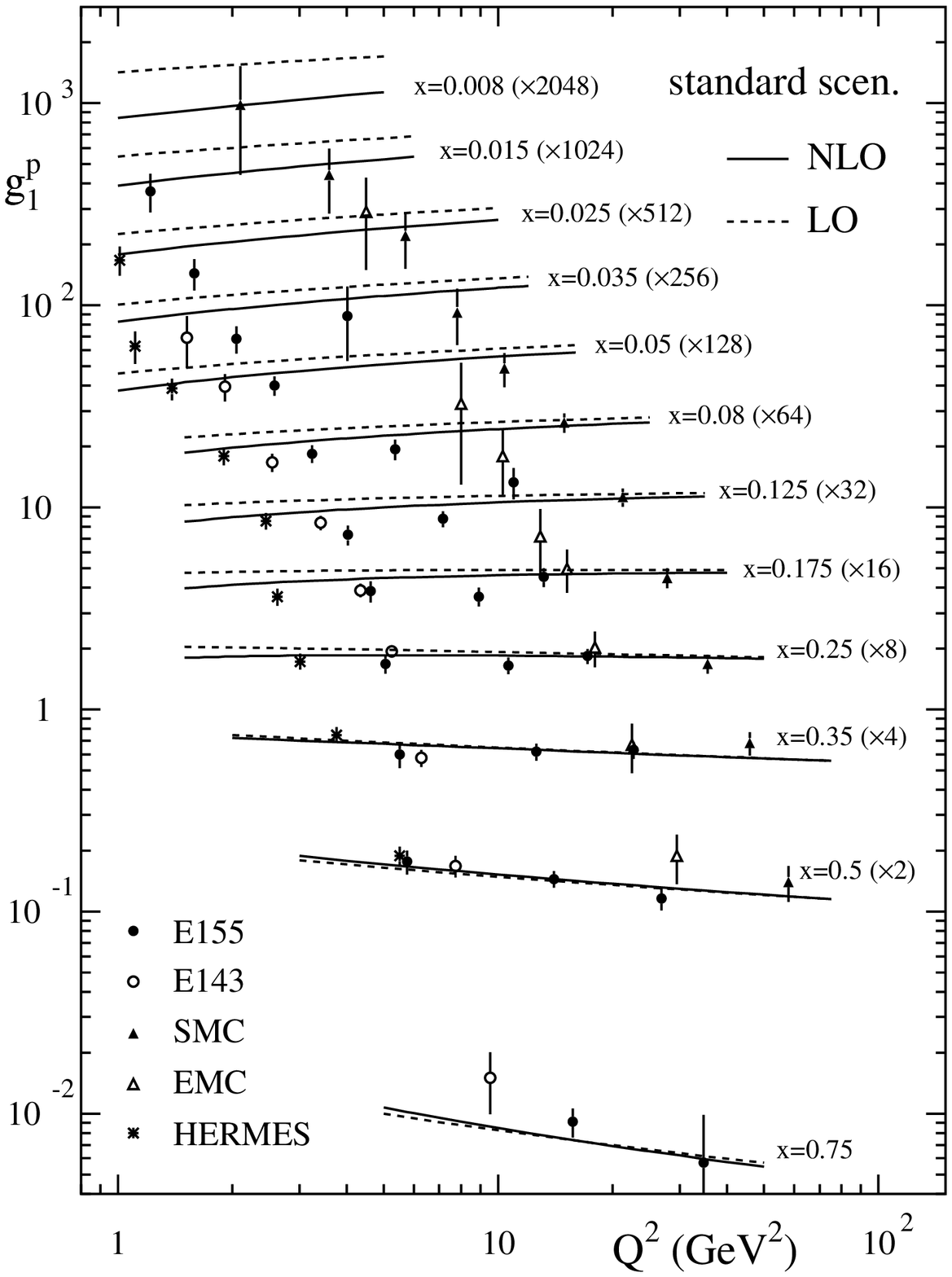,width=0.99\textwidth}
\vspace*{-2.0cm}
\large {\bf Fig.\ 2} \normalsize
\vspace*{-2.0cm}
\end{center}
\newpage

\vspace*{-2.0cm}
\begin{center}
\epsfig{file=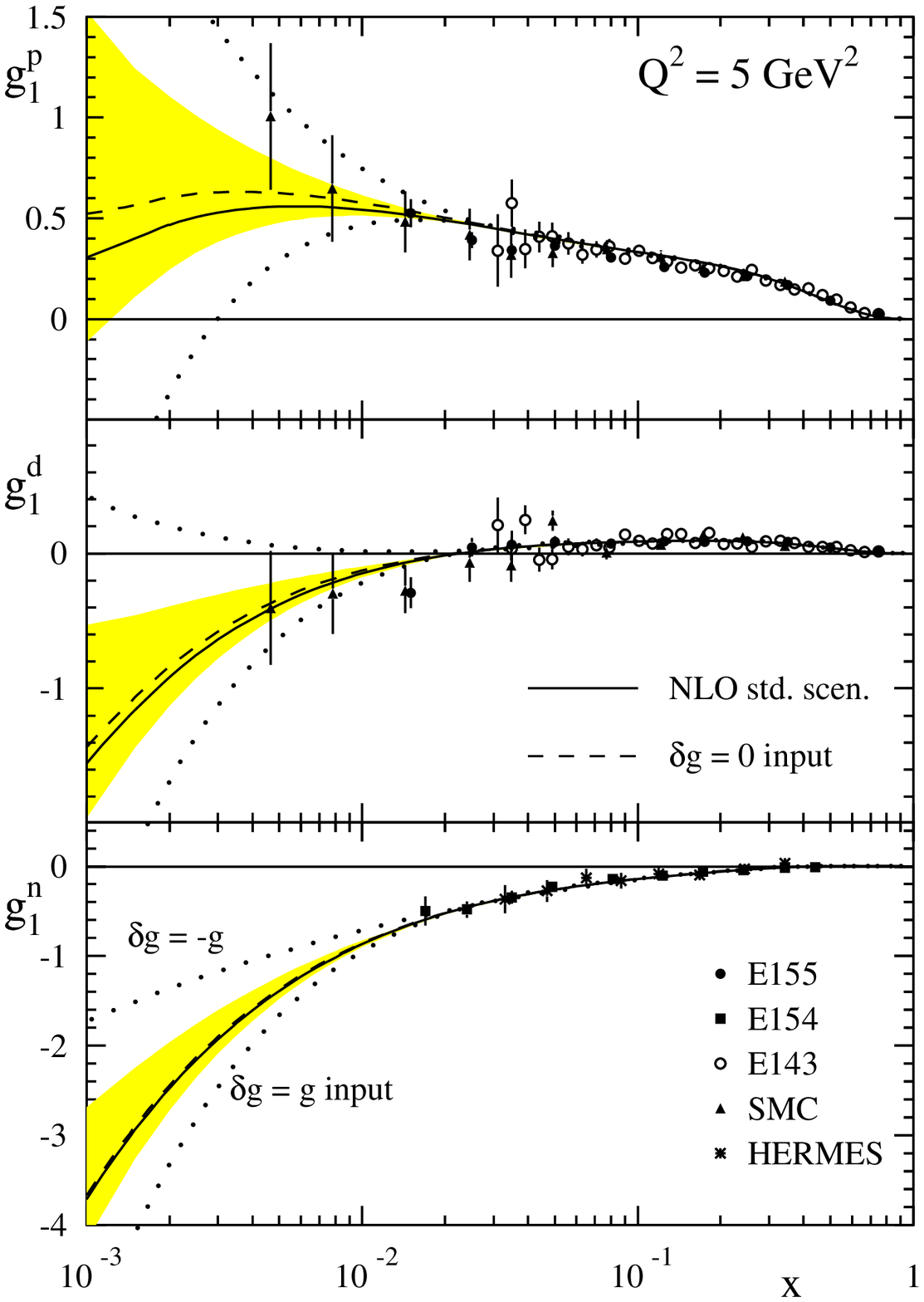,width=0.99\textwidth}

\vspace*{-0.2cm}
\large {\bf Fig.\ 3} \normalsize
\vspace*{-2.0cm}
\end{center}
\newpage

\hspace*{-1.8cm}
\begin{picture}(0,650)(0,0)
\epsfig{file=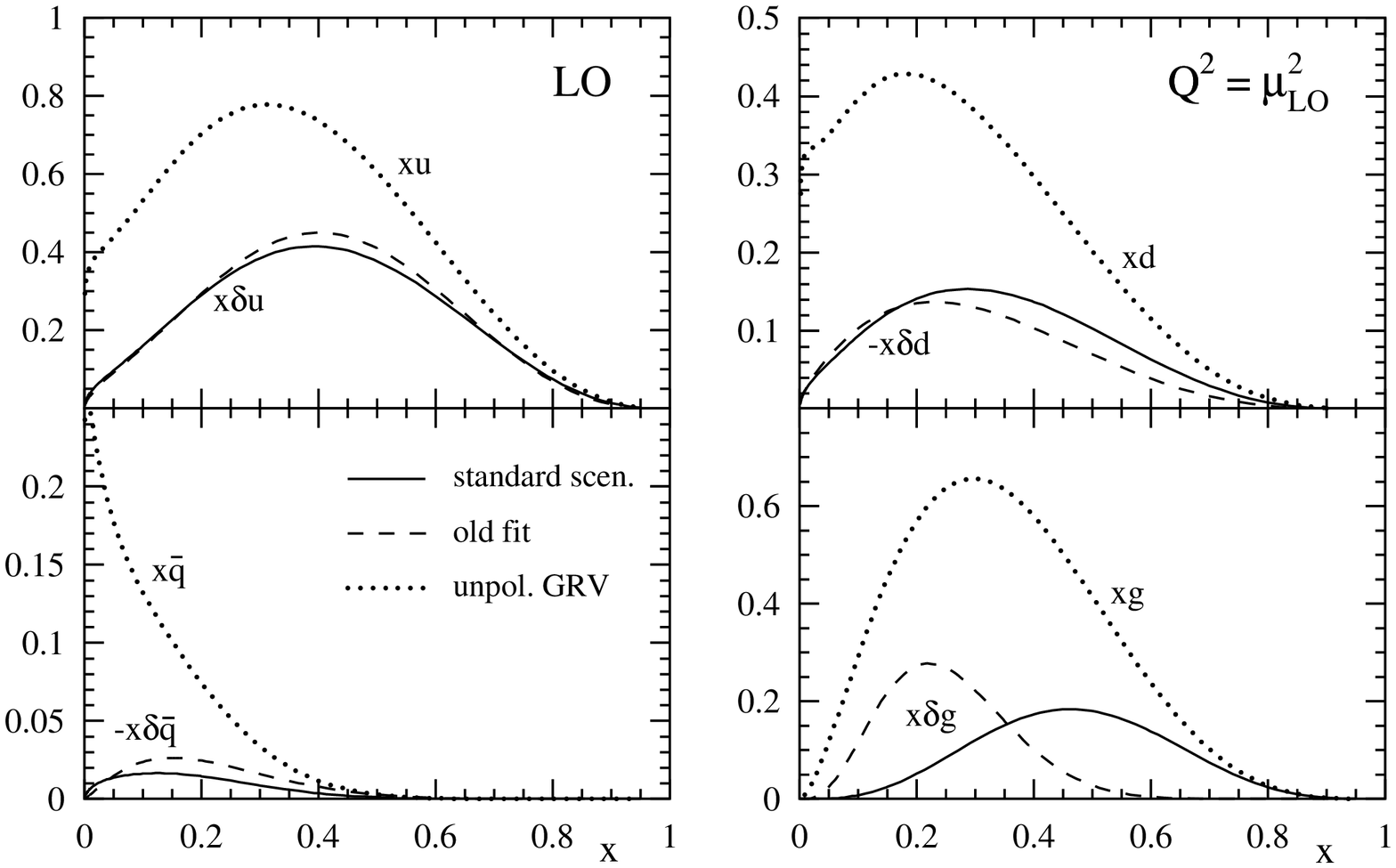,angle=90}
\put(10.,300.){\rotatebox{90}{\large {\bf Fig.\ 4(a)}}}
\end{picture}
\newpage

\hspace*{-1.8cm}
\begin{picture}(0,650)(0,0)
\epsfig{file=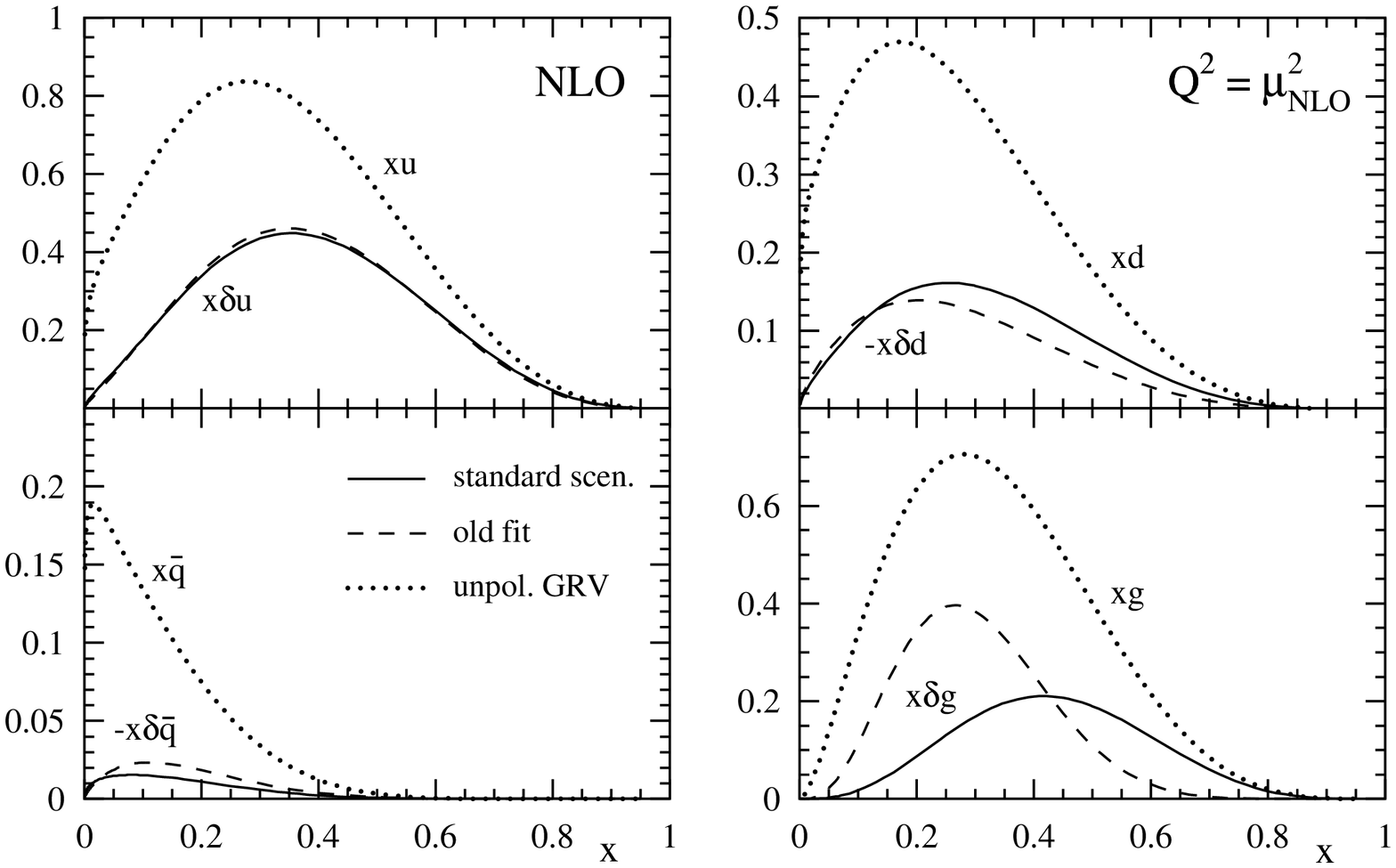,angle=90}
\put(10.,300.){\rotatebox{90}{\large {\bf Fig.\ 4(b)}}}
\end{picture}
\newpage

\vspace*{-2.0cm}
\hspace*{-1.8cm}
\begin{picture}(0,650)(0,0)
\epsfig{file=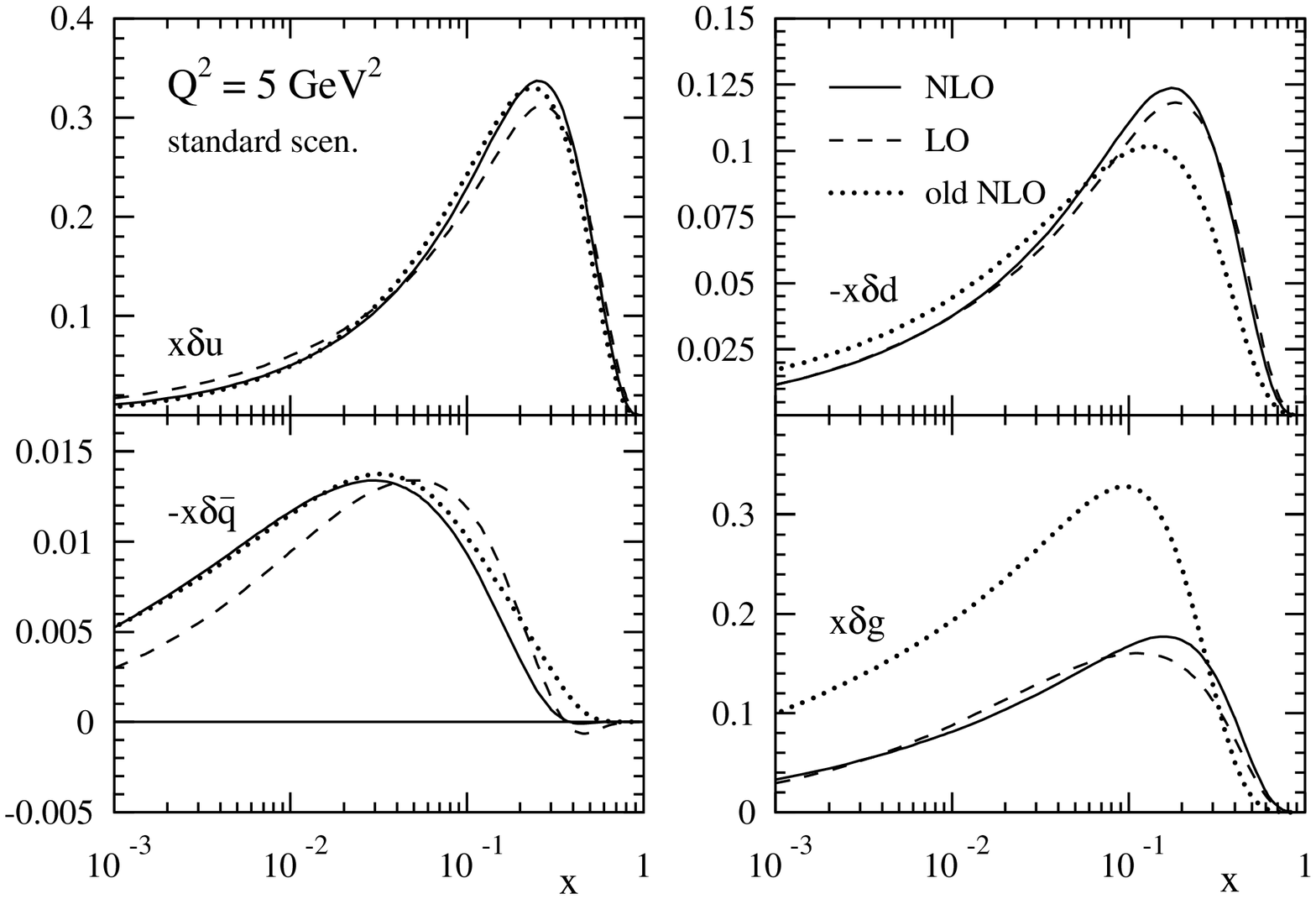,angle=90}
\put(10.,285.){\rotatebox{90}{\large {\bf Fig.\ 5}}}
\end{picture}
\newpage

\hspace*{-1.8cm}
\begin{picture}(0,650)(0,0)
\epsfig{file=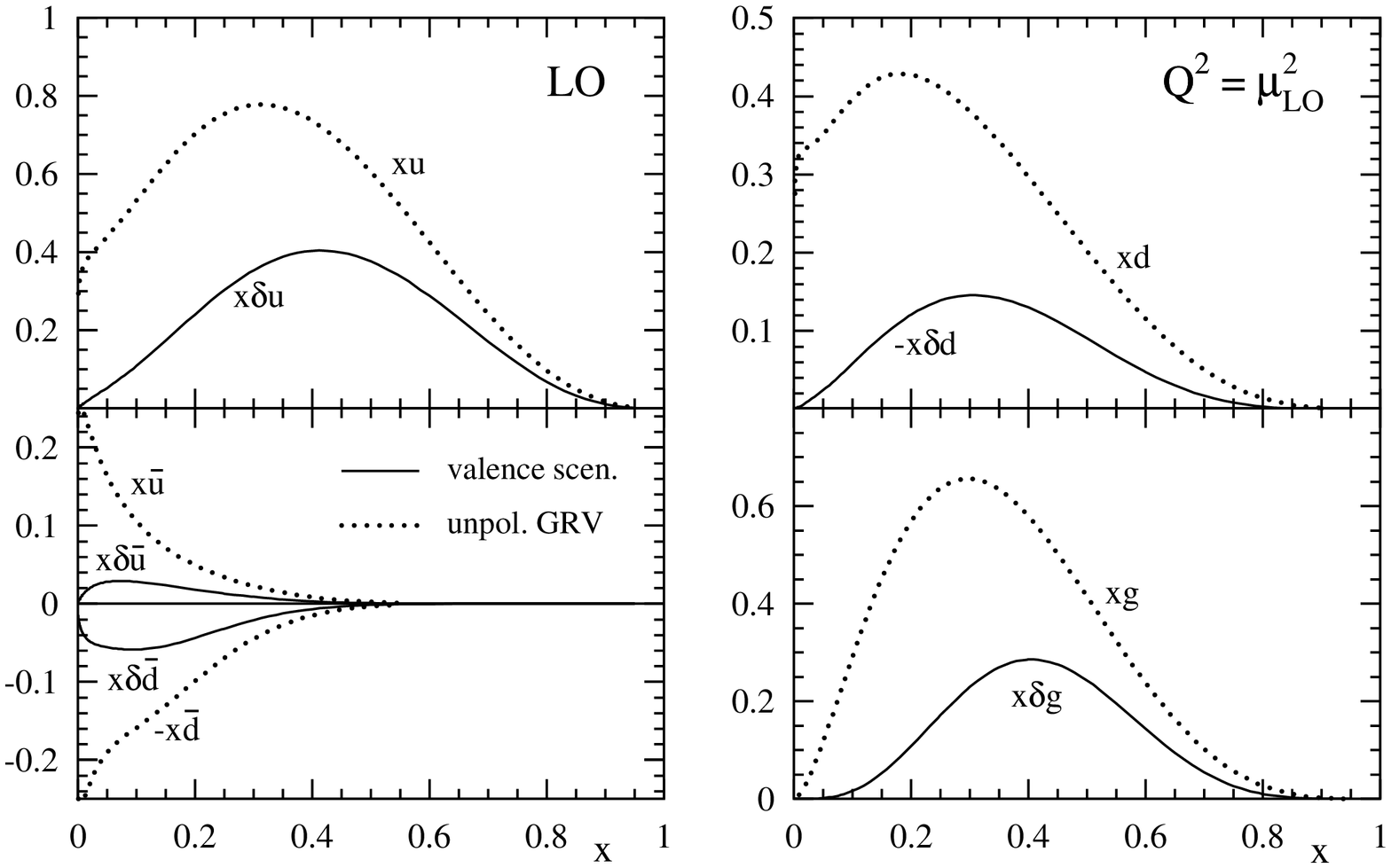,angle=90}
\put(10.,310.){\rotatebox{90}{\large {\bf Fig.\ 6(a)}}}
\end{picture}
\newpage

\hspace*{-1.8cm}
\begin{picture}(0,650)(0,0)
\epsfig{file=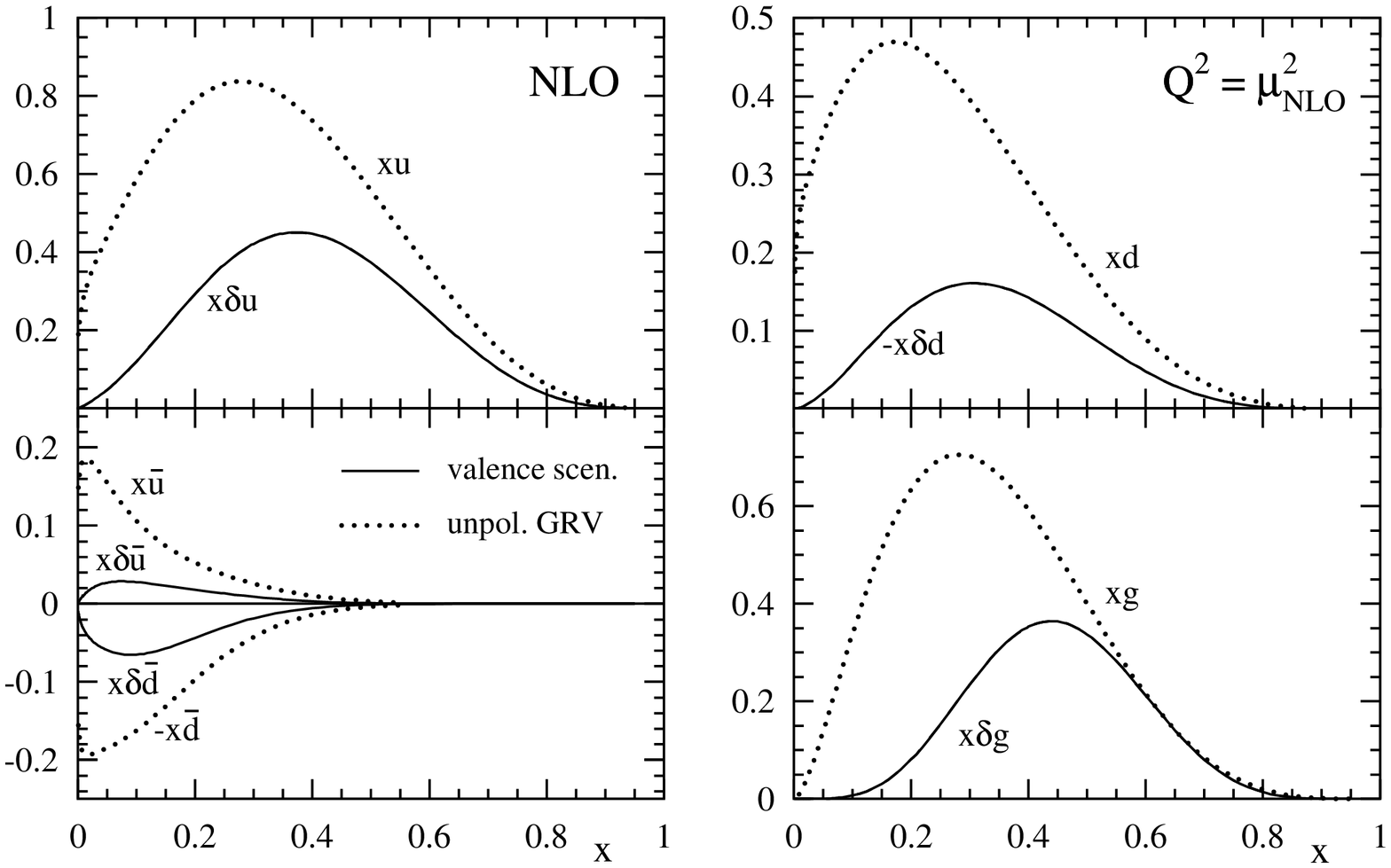,angle=90}
\put(10.,310.){\rotatebox{90}{\large {\bf Fig.\ 6(b)}}}
\end{picture}
\newpage

\vspace*{1.5cm}
\begin{center}
\epsfig{file=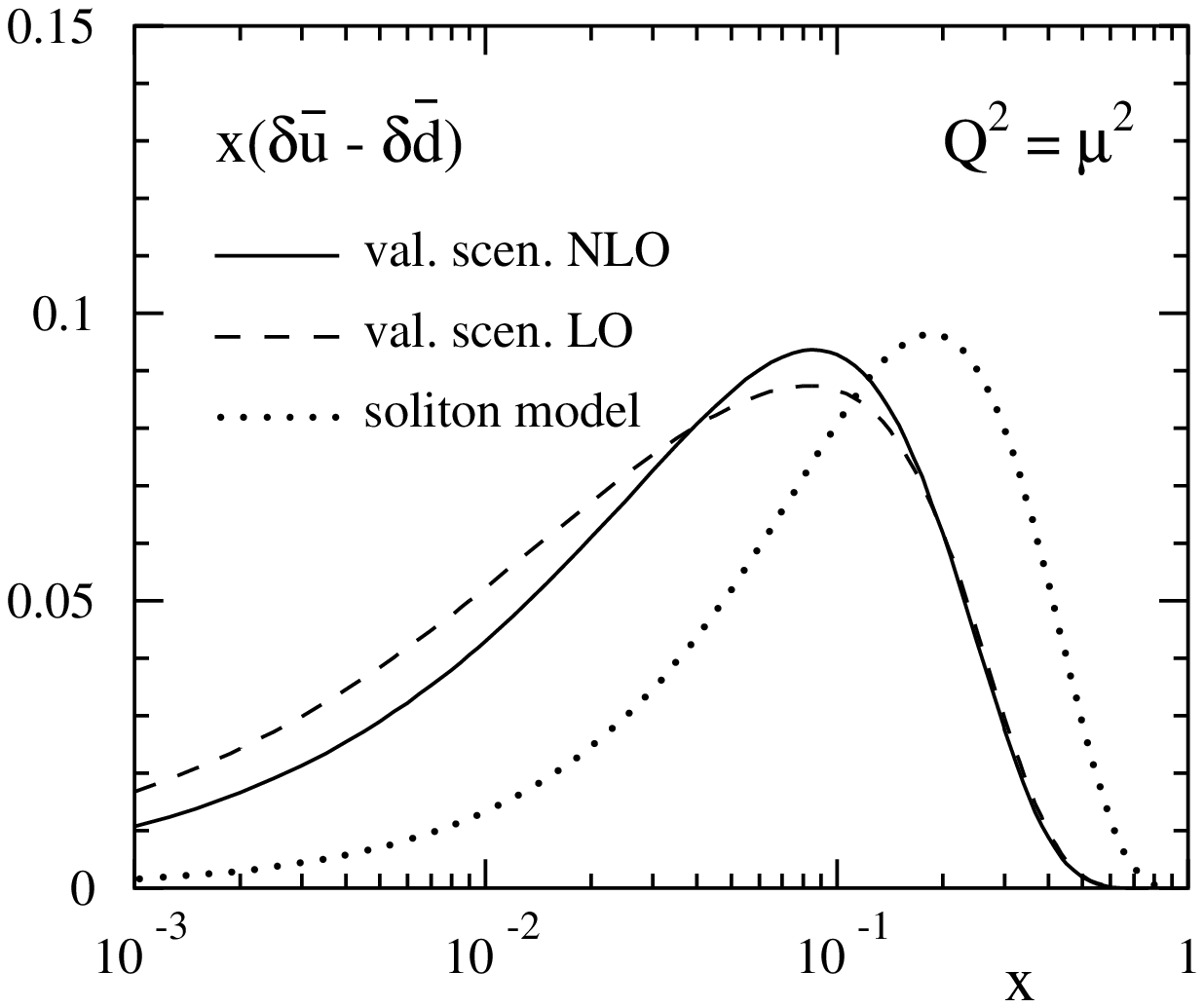,width=0.99\textwidth}
\large {\bf Fig.\ 7} \normalsize
\end{center}
\newpage

\vspace*{-2.5cm}
\hspace*{-1.0cm}
\begin{picture}(0,650)(0,0)
\epsfig{file=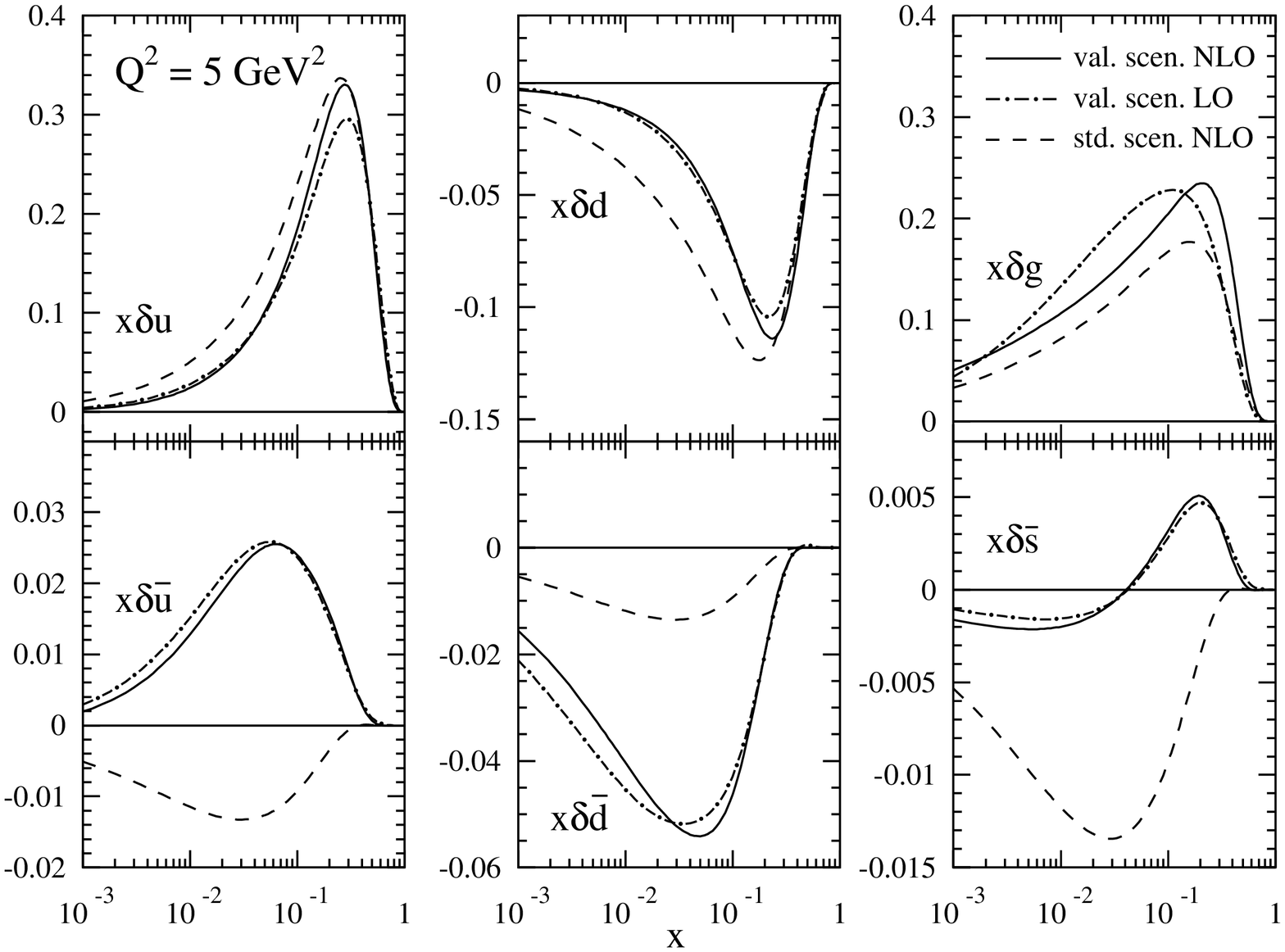,angle=90,width=0.99\textwidth}
\put(10.,290.){\rotatebox{90}{\large {\bf Fig.\ 8}}}
\end{picture}
\newpage

\vspace*{1.5cm}
\begin{center}
\epsfig{file=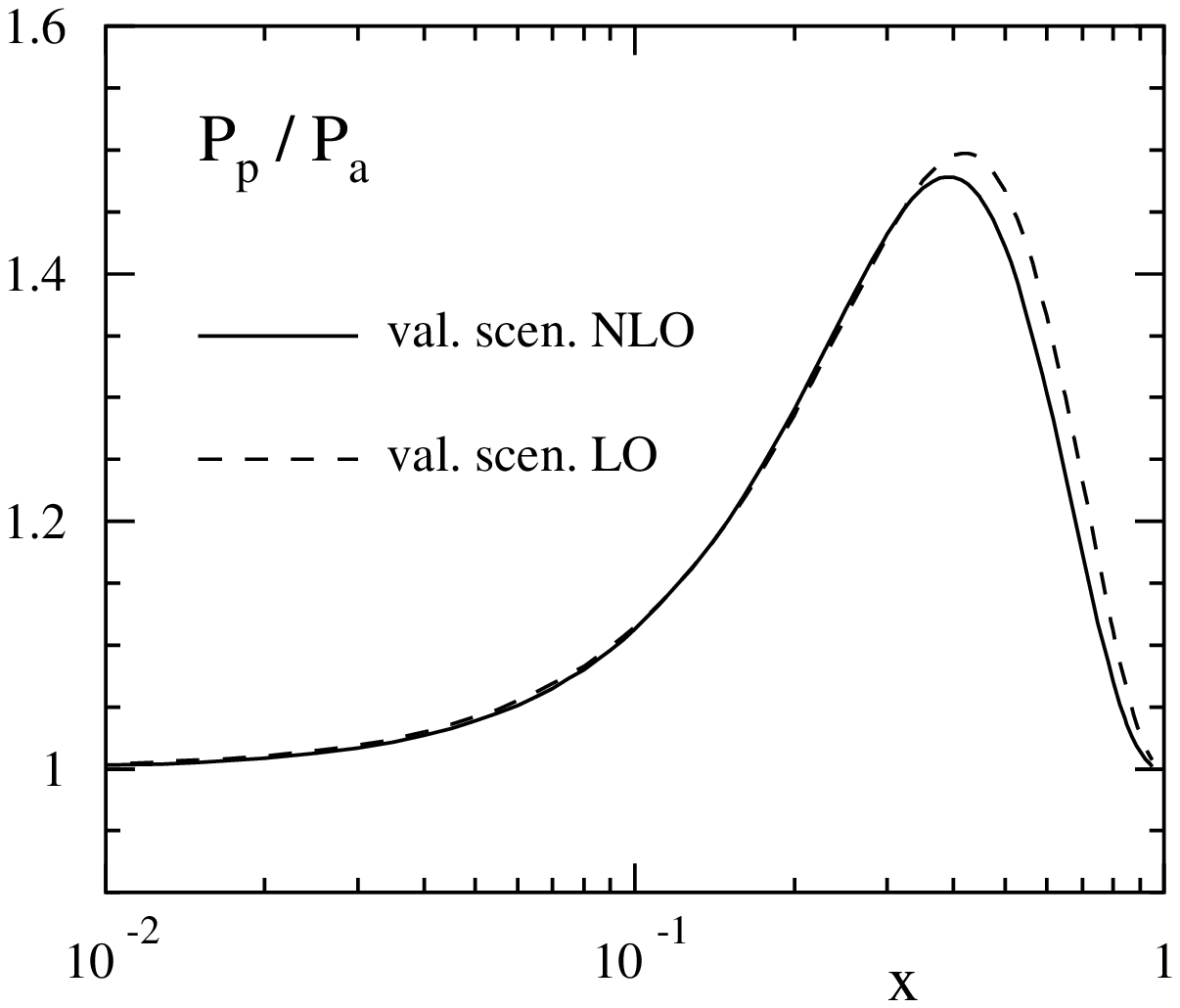,width=0.99\textwidth}
\large {\bf Fig.\ 9} \normalsize
\end{center}
\newpage

\vspace*{1.5cm}
\begin{center}
\epsfig{file=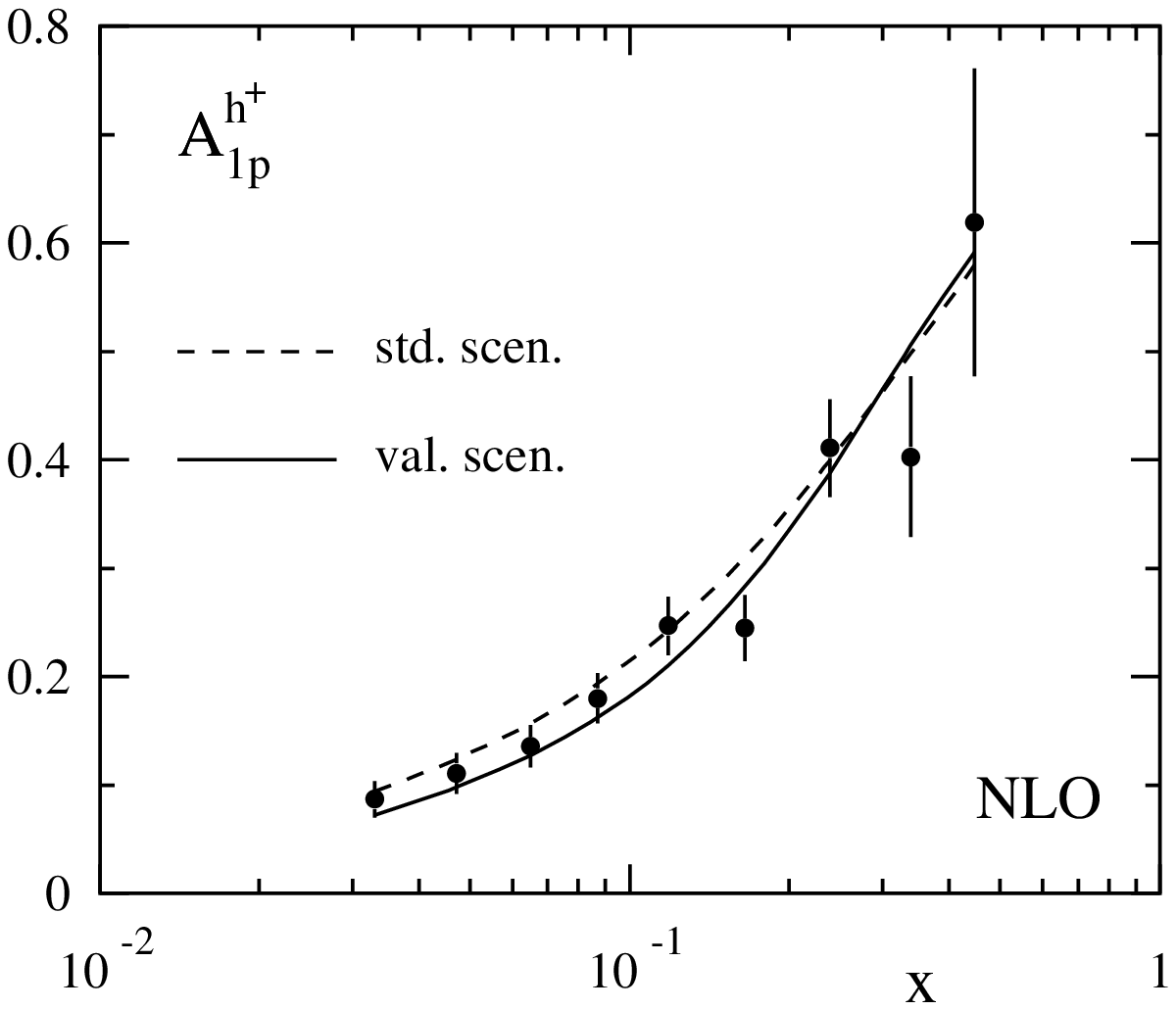,width=0.99\textwidth}
\large {\bf Fig.\ 10} \normalsize
\end{center}


\begin{thebibliography}{54}
\bibitem{ref37} Note that our ansatz violates (slightly) the
`helicity retention' concept developed in Ref.\ \cite{ref36}, by
which $\delta f(x,Q^2)/f(x,Q^2)\to 1$ for $x\to 1$.  Such
relations follow in perturbative QCD and need not necessarily be
fulfilled by our non--perturbative input at $Q^2=\mu^2$.  In any
case, we could always appropriately modify our ansatz at very
high $x$ so as to satisfy the `helicity retention' conditions.
Enforcing in our fits all $\beta_f$ to vanish in Eq.\ (2.6) increases
the total $\chi^2$ by several units, giving rise to worse fits than
our optimal ones.
\bibitem{ref51} The easiest and consistent way to implement
directly (even for Mellin--$n$ moments) the constraint (3.5) in a
fit routine is to fit, for example, $\delta u,\, \delta d$ and
$\delta\bar{u}$ according to the ansatz (2.6) which fix
$\delta\bar{d}(x,\mu^2)$ via (3.5) and (3.4):
$\delta\bar{d}/\bar{d} \simeq (\delta u/u)\,
(\delta\bar{u}/\bar{u})/(\delta d/d)$ where the unpolarized
densities are taken from Ref.\ \cite{ref25}.  Thus
$N_{\bar{d}}=N_u N_{\bar{u}}/N_d$, $\alpha_{\bar{d}} =
\alpha_u+\alpha_{\bar{u}}-\alpha_d$ and $\beta_{\bar{d}} =
\beta_u + \beta_{\bar{u}}-\beta_d$.
\end{thebibliography}
\end{document}